\documentclass[useAMS,usenatbib]{mn2e}
\usepackage{graphicx}
\usepackage{longtable}
\title[Quasar BAL variability measurements using un-absorbed
  reconstructions]{Quasar broad absorption line variability
  measurements using reconstructions of un-absorbed spectra}
\author[C. Wildy et al.]{C. Wildy$^{1}$\thanks{E-mail:
    cw268@le.ac.uk}, M. R. Goad$^{1}$ and
  J. T. Allen$^{2}$\\ $^{1}$University of Leicester, Department of
  Physics and Astronomy, University Road, Leicester, UK\\ $^{2}$Sydney
  Institute for Astronomy, School of Physics, A28, The University of
  Sydney, NSW 2006, Australia}

  \renewenvironment{thebibliography}[1]{%
    \begin{oldthebibliography}{#1}%
      \setlength{\parskip}{0ex}%
      \setlength{\itemsep}{0ex}%
  }%
  {%
    \end{oldthebibliography}%
  }

\begin{document}

\date{Accepted. Received}

\pagerange{\pageref{firstpage}--\pageref{lastpage}} \pubyear{0000}

\maketitle

\label{firstpage}

\begin{abstract}

\noindent We present a two-epoch Sloan Digital Sky Survey and Gemini/GMOS+William Herschel Telescope/ISIS variability study of 50 broad absorption line (BAL) quasars of redshift range $1.9 < z < 4.2$, containing 38 Si~{\sc iv} and 59 C~{\sc iv} BALs and spanning rest-frame time intervals of $\approx$10 months to 3.7 years. We find that 35/50 quasars exhibit one or more variable BALs, with 58\% of Si~{\sc iv} and 46\% of C~{\sc iv} BALs showing variability across the entire sample. On average, Si~{\sc iv} BALs show larger fractional change in BAL pseudo equivalent width than C~{\sc iv} BALs, as referenced to an unabsorbed continuum+emission-line spectrum constructed using non-negative matrix factorisation. No correlation is found between BAL variability and quasar luminosity, suggesting that ionizing continuum changes do not play a significant role in BAL variability (assuming the gas is in photoionization equilibrium with the ionizing continuum). A subset of 14 quasars have one variable BAL from each of Si~{\sc iv} and C~{\sc iv} with significant overlap in velocity space and for which variations are in the same sense (strengthening or weakening) and which appear to be correlated (98\% confidence). We find examples of both appearing and disappearing BALs in weaker/shallower lines with disappearance rates of 2.3\% for C~{\sc iv} and 5.3\% for Si~{\sc iv}, suggesting average lifetimes of 142 and 43 years respectively. We identify 5 objects in which the BAL is coincident with the broad emission-line, but appears to cover only the continuum source. Assuming a clumpy inhomogeneous absorber model and a typical size for the continuum source, we infer a maximum cloud radius of 10$^{13}$ to 10$^{14}$~cm, assuming Eddington limited accretion.

\end{abstract}


\section{Introduction}

Broad absorption line quasars (BALQSOs) show evidence of high-velocity outflows in the form of strong, blue-shifted broad absorption lines (BALs). Previous quasar spectral studies have indicated that this sub-category comprises $10\%-20\%$ of the QSO population \citep{reichard03a,knigge08,scaringi09}, although the intrinsic fraction may be as high as 41$\%$ if differential SDSS target
selection effects are taken into account \citep{allen11}.

The BALs, by definition, comprise absorption troughs in which the flux is less than 90$\%$ of the continuum level, extending over at least 2000 km s$^{-1}$ in the quasar rest frame \citep{weymann91}. They are one of three examples of blue shifted absorption features seen in AGN, others being NALs, with velocity widths of a few hundred km s$^{-1}$, and mini-BALs, with velocity widths intermediate to those of NALs and BALs \citep{narayanan04}. The outflows giving rise to these absorption lines are thought to originate in accretion disc instabilities and are accelerated to velocities of up to 0.2c in the quasar rest frame \citep{hamann13} by radiation pressure \citep{elvis00,proga00,chelouche01}, magnetic effects \citep{blandford82,everett05} or thermal winds \citep{bottorff97,giustini12} from the accretion disc. The fraction of quasars presenting BALs can be interpreted as an orientation effect, with this fraction being a measure of the proportion of the sky as seen from the quasar which is covered by the outflow. Large-scale outflows may therefore occur in a much greater fraction of the quasar population than the BAL fraction would imply \citep{schmidt99}.

The flows have been theorised to extend as far as a few hundred parsecs from the central engine \citep{barlow94}, allowing the kinetic energy to be transferred to the interstellar medium. This process may be responsible for quenching star formation in the host galaxy and limiting black hole growth \citep{springel05,king10}. It may also result in the observed relationship between the masses of supermassive
black holes and galactic bulges \citep{magorrian98,silk98}.

The most common ($\sim$85 per cent) BALQSO subtype is the High-Ionization BALQSOs (HiBALs) whose spectra only show broad absorption due to high ionization transitions such as C~{\sc iv} $\lambda$1549, Si~{\sc iv} $\lambda$1400 and N {\sc v} $\lambda$1240 \citep{sprayberry92,reichard03b}. The remaining proportion, known as Low Ionization BALQSOs (LoBALs) show broad absorption resulting from low ionization species such as Mg~{\sc ii}, Al~{\sc iii} and on rare occasions Fe~{\sc ii} (known as FeLoBALs) in addition to high ionization BALs.

Several studies now exist which probe the spectral variability of broad absorption lines in BALQSOs over a range of object rest-frame timescales from approximately 1 week to several years \citep{barlow94,lundgren07,gibson08,gibson10,capellupo11,capellupo12,capellupo13}. Measurements of quasar variability can provide valuable insights into the properties of the central engine. It is known that photoionization effects resulting from changes in the ionizing continuum drive the observed broad emission line (BEL) variability \citep{peterson98,vandenberk04,wilhite06}, however the mechanism responsible for BAL variations is less secure. Variability studies of broad absorption lines can prove important in identifying the geometry of the outflows as well as constraining the range of physical conditions that result in variability, especially since previous results have proved contradictory or inconclusive. Evidence has been put forward in favour of ionization state changes \citep{barlow94}, movement of gas along the line of sight to the quasar emission region \citep{gibson08,hamann08}, or a combination of both effects \citep{capellupo12}.

Many previous studies have focused on C~{\sc iv} BAL variability only \citep{lundgren07,gibson08,capellupo11,filizak12}, however recent efforts have also included significant Si~{\sc iv} investigations \citep{gibson10,capellupo12}. We examine variability in both ions from a sample consisting mostly of HiBAL QSOs. Of the previous investigations into C~{\sc iv} BALs, \citet{lundgren07} (hereafter referred to as L07) and \citet{filizak12} used minimum velocity cutoffs to avoid the difficulties associated with properly identifying and quantifying absorption where broad emission lines (BELs) overlap with the corresponding absorption line. Many BALQSOs show evidence of absorption of both continuum and emission flux by the BAL outflow \citep{turnshek88}, meaning that in cases of BAL/BEL overlap, more light is removed from the line of sight than a continuum based BAL depth measurement would suggest. We address these issues by attempting to quantify variability in BALs which include features down to zero velocity using reconstructions of the un-absorbed quasar spectrum over the wavelength range of interest. As well as solving the problem of introducing a low velocity cutoff, this allows measurements to be taken for cases of high velocity ($>$13 500 km s$^{-1}$) Si~{\sc iv} outflows, which may overlap with the O {\sc i} $\lambda$1304 and C~{\sc ii} $\lambda$1334 emission lines \citep{trump06}. Using the method of \citet{allen11}, we produce reconstructions of the underlying continuum plus emission spectra for each of the quasars in our sample using non-negative matrix factorisation (NMF), a blind source separation technique. We use this un-absorbed spectrum as a pseudo continuum from which BAL pseudo equivalent width measurements are made.

The ionization energies of Si {\sc iii} and Si~{\sc iv} are 33.5 eV and 45.1 eV respectively, while the ionization energies of C {\sc iii} and C~{\sc iv} are 47.9 eV and 64.5 eV. The abundances of carbon and silicon are also assumed to be unequal \citep{hamann11}. Differing variability properties of Si~{\sc iv} and C~{\sc iv} can therefore inform attempts to explain the physics, kinematics and geometry of
outflows which result in BAL variability. A significant fraction (41 out of 50 quasars) of our sample contain both Si~{\sc iv} and C~{\sc iv} BALs, allowing direct comparison in the behaviour of the two
ions in these quasars.

Throughout this paper all time intervals are measured in days in the quasar rest frame. We assume a flat $\Lambda$CDM cosmology with H$_{0}$=70 km s$^{-1}$Mpc$^{-1}$,$\Omega$$_{M}$=0.3 and
$\Omega$$_{\Lambda}$=0.7.

\section{Sample selection and observations}

	  \subsection{The Balnicity Index}

A common method used to identify a quasar as a BALQSO is to consider a quantity known as the Balnicity Index (BI), originally defined by \citet{weymann91} for C~{\sc iv} BALs. The BI is a modified equivalent width (EW) measurement which is zero for non-BALQSOs and positive for BALQSOs. The BI is defined as

\begin{equation}
\label{eqn:bi}
B.I.=\int_{-25000}^{-3000} \left[1-\frac{f(v)}{0.9}\right]Cdv
\end{equation}

\noindent where $f(v)$ is the continuum normalised flux as a function of outflow velocity $v$ relative to line centre and C is a constant equal to unity when $f(v)$ has been under 90$\%$ of the continuum
level over the previous 2000 km s$^{-1}$ of the integration range and zero otherwise. The upper velocity boundary avoids confusion between C~{\sc iv} outflows and the Si~{\sc iv} emission line. The lower
velocity boundary avoids absorption line overlap with the transition's broad emission line.  An alternative to the BI is the absorption index (AI) defined in \citet{hall02}. The AI modifies the BI so that it requires at least a contiguous 450 km s$^{-1}$ of absorption at below the 90$\%$ continuum level to register a positive value rather than 2000 km s$^{-1}$ and reduces the lower velocity limit to 0 km
s$^{-1}$.

	\subsection{Sample selection}

A redshift range of approximately 1.9$<z<$4.1 was selected so that the spectral region from the continuum band at 1350\AA{} to the continuum red-ward of C~{\sc iv} $\lambda$1549 was
observable. Variability is tested across two epochs, hereafter referred to as epoch 1 and epoch 2 for the earlier and later observations respectively, covering a rest frame timescale of $\approx$10 months to 3.7 years. This epoch separation timescale is relatively unexplored in previous BAL variability studies. Epoch 1 uses spectral data from objects imaged as part of the Sloan Digital Sky Survey, while epoch 2 observations were obtained from either the William Herschel Telescope
or the Gemini Observatories.

The BALQSOs observed as part of this study were selected from the BALQSO catalogue of \citet{scaringi09}. This catalogue comprises 3552 BALQSOs and was compiled from a parent sample of 28421 QSOs published
in the 5th data release (DR5) of the Sloan Digital Sky Survey \citep{adelmanmccarthy07} using a hybrid classification scheme comprising a combination of simple metrics (e.g Balnicity Index), a supervised neural network and visual inspection.

From the parent sample of 3552 BALQSOs we select only those objects tagged as definitely real (1353), and with BI$> 2,000$ km/s to minimise contamination by (possibly unrelated) narrow absorption line
systems. A total of 595 objects satisfy all of these criteria. From this sample, objects were selected to span a broad range in balnicity ($BI> 2000$km/s), redshift ($z>2$ ) and luminosity ($-28.75 < M_{i} < -25.75$). Of the 50 objects from this sample observed as part of this program, a K-S test on the distribution functions of their observed properties ($BI$, $z$, $M_{i}$), shows that they are statistically
consistent with those of the parent BALQSO population.

Non-BALQSO continuum variability is known to scale inversely with object luminosity and directly with redshift \citep{vandenberk04}. As BALQSOs seem to be part of the same parent population as non-BALQSOs
\citep{reichard03b} and given that ionization is the main driver of variability in non-BALQSOs \citep{blandford82,peterson98}, if ionizing continuum variations produce the variability seen in BALs it may be
expected that lower luminosity BALQSOs would show the greatest BAL variations. In order to test the relationship between luminosity and variability our BALQSO sample spans a factor of 10 in i-band
source luminosity.  The redshift distribution of the BALQSOs from our original sample observed as part of this campaign peaks at the lower limit of our sample redshift range and is too small to give
statistically reliable measurements of the dependence of BAL variability on redshift.

	  \subsection{The Sloan Digital Sky Survey}

Epoch 1 observations were obtained from DR6 of the Sloan Digital Sky Survey (SDSS). Beginning in 2000, the SDSS had imaged approximately 10000 deg$\sp{2}$ of the sky as of June 2010 \citep{schneider10},
using the 2.5m Telescope at the Apache Point Observatory, New Mexico, USA \citep{sdss}. Imaging is carried out using a CCD camera that operates in drift-scanning mode \citep{gunn98}. Five broad band
filters, u,g,r,i and z, normalised to the AB system, are used, covering a wavelength range of 3900 to 9100 \AA{} with an approximate spectral resolution of $\lambda$/$\Delta$$\lambda$=2000 at 5000 \AA{}
\citep{stoughton02}. Further information on the SDSS data reduction process is given in \citet{lupton01} and \citet{stoughton02}.

          \subsection{Gemini North and South Observatories}

Gemini-North/GMOS and Gemini-South/GMOS are spectrographs consisting of three CCD detectors separated by gaps corresponding to 39 unbinned pixels, which were used to obtain data for 44 target observations in long-slit mode. Five of these observations were excluded from the final sample due to cloud cover issues, insufficient integration time or a deficiency of spectral regions containing unabsorbed continuum.

A slit width of 1.5" was used at both locations for all Gemini observations. The grating configurations employed blaze wavelengths such that the spectral region of interest, extending from 1350\AA{} to
C~{\sc iv} $\lambda$1549 in the quasar rest frame, was included in the total wavelength span. The B600 grating (resolution $\Delta$$\lambda$/$\lambda$$\approx$1688 at blaze wavelength 4610\AA{}) and R400 grating (resolution $\Delta$$\lambda$/$\lambda$$\approx$1918 at blaze wavelength 7640\AA{}) were used. Three were observed in semester A 2008, the rest throughout 2011 and early 2012. Details of Gemini/GMOS observations are given in Table 1 (lower panel).

The GMOS spectra were reduced using Version 1.11 of the Gemini IRAF package, initially applying bias correction using zero time exposures and using flat field images to correct for illumination differences
and pixel to pixel sensitivity variations in the CCD images. Spectra were extracted using the IRAF optimal extraction routine using a width of 10 pixels. Wavelength calibration and flux calibration were applied using arc line exposures and observations of photometric standard stars respectively. Arc exposures and standard star observations were obtained during the observing nights. Sky background subtraction was also applied to the extracted spectra to remove much of the atmospheric background contaminating the data. Details of eachg Gemini object observation are given in Table 1 (lower panel.)

	  \subsection{The William Herschel Telescope}

The William Herschel Telescope (WHT) ISIS spectrograph was used for longslit spectroscopy of eight targets during semester A 2008, taken as part of a previous investigation into radiative line driving
\citep{cottis10}. Observations were taken with the ISIS double-armed (red and blue arm) spectrograph using the 5700 dichroic allowing simultaneous observations at both short and long wavelengths. For each
object, spectra from the two arms in the overlapping range 5400 to 5700 \AA{} were combined using an error weighted mean in each wavelength bin. Observations were performed in long-slit mode with the
R158B and R158R gratings, giving a nominal spectral resolution of 1.6\AA{} per pixel in the blue and 1.8\AA{} per pixel in the red. A slit width of 1.5" was chosen to match typical seeing conditions and
give reasonable throughput without compromising the spectral resolution. For each source, multiple exposures were taken (in order to remove cosmic ray hits) bracketed by arc-lamp exposures for wavelength calibration. Standard star spectra were taken on the observing nights to allow flux calibration of the targets.  As in the Gemini/GMOS observations, correction of CCD images using bias and flat field exposures was performed, this time within the IRAF \emph{ccdproc} routine. Spectra were extracted, along with sky background removal, using the \emph{apall} task within the IRAF \emph{longslit} package. Wavelength and flux calibration were applied to the extracted spectra within this same package.

Details of each WHT object observation are given in Table 1 (upper panel). 

\begin{table*}
\begin{center}
\caption{WHT/ISIS (upper panel) and Gemini/GMOS (lower panel) long-slit spectroscopic observations of SDSS quasars.}
\begin{tabular}{l l l l l l}
\hline Object&z&Absolute i-band Magnitude&Exp. time&Grating&Mean Airmass\\
 & &M$_{i}$&(s)& & \\
\hline
SDSSJ101844.45+544015.6$^{d}$&3.251$\pm$0.004&-26.852&3$\times$1500&R158B $\&$ R158R&1.26\\
SDSSJ113831.42+351725.3$^{d}$&2.122$\pm$0.002&-26.891&3$\times$1200&R158B $\&$ R158R&1.14\\
SDSSJ114704.47+153243.3$^{d}$&3.092$\pm$0.002&-27.848&2$\times$1200&R158B $\&$ R158R&1.29\\
SDSSJ134458.82+483457.5$^{d}$&2.048$\pm$0.002&-26.709&3$\times$1500&R158B $\&$ R158R&1.24\\
SDSSJ162657.47+405848.0$^{d}$&3.062$\pm$0.003&-27.576&4$\times$1400&R158B $\&$ R158R&1.04\\
SDSSJ164152.30+305851.7$^{d}$&2.016$\pm$0.002&-27.831&3$\times$1500&R158B $\&$ R158R&1.06\\
SDSSJ170056.85+602639.7$^{d}$&2.136$\pm$0.002&-27.367&3$\times$1400&R158B $\&$ R158R&1.25\\
SDSSJ212412.60+095923.3$^{d}$&1.921$\pm$0.002&-26.665&3$\times$1200&R158B $\&$ R158R&1.17\\
\hline
SDSSJ001025.90+005447.6&2.861$\pm$0.002&-27.806&2$\times$600&B600+G5323&1.54\\
SDSSJ004613.54+010425.7&2.149$\pm$0.002&-28.404&2$\times$600&B600+G5323&1.32\\
SDSSJ025720.43-080322.5&2.045$\pm$0.005&-26.786&2$\times$1500&B600+G5323&1.37\\
SDSSJ031033.45-060957.8&2.050$\pm$0.002&-27.424&2$\times$900&B600+G5323&1.26\\
SDSSJ031331.22-070422.8&2.777$\pm$0.002&-27.844&2$\times$1200&B600+G5323&1.51\\
SDSSJ032832.77-070750.3$^{a}$&2.917$\pm$0.005&-27.223&2$\times$1500&B600+G5323&1.36\\
SDSSJ033223.51-065450.5&3.708$\pm$0.008&-27.646&1$\times$1500&B600+G5323&1.10\\
SDSSJ033224.95-062116.1&2.761$\pm$0.002&-27.487&2$\times$900&B600+G5323&1.50\\
SDSSJ034946.61-065730.3$^{b}$&3.973$\pm$0.011&-27.313&2$\times$1800&B600+G5323&1.27\\
SDSSJ035335.67-061802.5&2.165$\pm$0.002&-27.218&2$\times$1200&B600+G5323&1.67\\
SDSSJ035749.11-061121.9&2.006$\pm$0.002&-26.976&2$\times$1400&B600+G5323&1.53\\
SDSSJ073535.44+374450.4&2.751$\pm$0.002&-28.259&1$\times$1500&B600+G5307&1.10\\
SDSSJ081823.46+484910.8&2.015$\pm$0.002&-26.584&4$\times$900&B600+G5307&1.28\\
SDSSJ081925.00+032455.7&2.239$\pm$0.002&-27.022&2$\times$600&B600+G5323&1.36\\
SDSSJ082813.47+065326.4&2.968$\pm$0.004&-27.779&2$\times$900&B600+G5323&1.35\\
SDSSJ083718.63+482806.1&3.646$\pm$0.003&-28.484&1$\times$1200&B600+G5307&1.25\\
SDSSJ084023.51+063739.1$^{c}$&3.801$\pm$0.010&-27.170&1$\times$1362&B600+G5323&1.25\\
SDSSJ084408.29+423226.9&2.964$\pm$0.002&-28.215&1$\times$1500&B600+G5307&1.40\\
SDSSJ085006.08+072959.0&2.690$\pm$0.002&-28.622&2$\times$300&B600+G5323&1.41\\
SDSSJ085104.05+051539.8$^{d}$&3.222$\pm$0.003&-28.480&6$\times$600&B600+G5303&1.04\\
SDSSJ092557.52+044035.9&2.271$\pm$0.002&-27.174&2$\times$600&B600+G5323&1.53\\
SDSSJ092639.34+383656.7&2.155$\pm$0.002&-27.292&2$\times$1200&B600+G5307&1.32\\
SDSSJ093251.98+023727.0&2.169$\pm$0.002&-28.106&2$\times$400&B600+G5323&1.37\\
SDSSJ095224.84+064732.0&2.174$\pm$0.002&-27.877&2$\times$400&B600+G5323&1.42\\
SDSSJ100021.72+035116.5&2.017$\pm$0.002&-27.314&1$\times$600&B600+G5323&1.32\\
SDSSJ100312.63+402505.6&3.247$\pm$0.003&-27.342&3$\times$900&B600+G5307&1.38\\
SDSSJ101056.68+355833.3&2.301$\pm$0.002&-27.334&1$\times$1200&B600+G5307&1.04\\
SDSSJ104059.79+055524.4&2.450$\pm$0.002&-27.023&2$\times$1500&B600+G5323&1.40\\
SDSSJ105334.57+425724.9&2.719$\pm$0.003&-27.221&4$\times$900&B600+G5307&1.17\\
SDSSJ110041.19+003631.9&2.020$\pm$0.002&-27.497&2$\times$600&B600+G5323&1.22\\
SDSSJ110339.90+011928.5&2.056$\pm$0.002&-27.203&2$\times$600&B600+G5323&1.28\\
SDSSJ111516.08+460234.6$^{c}$&4.175$\pm$0.010&-27.376&1$\times$721 $\&$ 1$\times$900&R400+G5305&1.16\\
SDSSJ112733.69+343008.8&4.060$\pm$0.009&-28.594&4$\times$900&R400+G5305&1.06\\
SDSSJ114722.09+373720.7&2.199$\pm$0.002&-26.380&4$\times$900&B600+G5307&1.05\\
SDSSJ115007.66+542737.1$^{d}$&3.534$\pm$0.004&-28.016&6$\times$600&R400+G5305&1.22\\
SDSSJ125628.67+393548.0&2.138$\pm$0.002&-27.309&1$\times$1200&B600+G5307&1.07\\
SDSSJ142244.45+382330.6$^{d}$&3.741$\pm$0.010&-28.566&6$\times$600&R400+G5305&1.06\\
SDSSJ143604.64+350428.5&3.035$\pm$0.006&-26.625&3$\times$900&B600+G5307&1.04\\
SDSSJ143632.25+501403.6&2.784$\pm$0.002&-27.502&2$\times$1500&B600+G5307&1.67\\
SDSSJ151601.51+430931.4&2.610$\pm$0.002&-28.373&2$\times$1200&B600+G5307&1.11\\
SDSSJ153226.22+313138.1$^{c}$&2.889$\pm$0.003&-28.194&1$\times$336 $\&$ 1$\times$1200&B600+G5307&1.21\\
SDSSJ165248.29+325032.3&2.832$\pm$0.008&-26.558&1$\times$1500&B600+G5307&1.03\\
SDSSJ205659.48-071123.1&2.083$\pm$0.002&-27.547&2$\times$900&B600+G5323&1.10\\
SDSSJ210436.62-070738.3&2.360$\pm$0.004&-27.268&2$\times$1500&B600+G5323&1.17\\
SDSSJ211718.17+010248.9&2.928$\pm$0.003&-27.550&2$\times$1500&B600+G5323&1.17\\
SDSSJ213138.93-070013.3&2.045$\pm$0.002&-27.425&2$\times$900&B600+G5323&1.37\\
SDSSJ222505.28-084542.7&2.085$\pm$0.003&-27.121&2$\times$900&B600+G5323&1.17\\
\hline    
\end{tabular}
\label{tab:obs}
\end{center}
$^{a}$ Excluded due to lack of observable continuum bands $<$1549\AA{}\\
$^{b}$ Excluded due to 80$\%$ cloud cover\\
$^{c}$ Excluded due to shortened integration time.\\
$^{d}$ Observed in 2008.\\
Redshifts taken from \citet{hewett10}
\end{table*}

\section{Reconstruction Method}

The unabsorbed continuum plus emission line spectra -- the pseudo continuum spectra -- were reconstructed using the non-negative matrix factorisation (NMF) method described in \citet{allen11}. For the epoch
1 (SDSS) spectra the reconstructions were the same as those used in \citet{allen11}; new fits were produced for the epoch 2 (Gemini and WHT) spectra, with some minor modifications to the method. Full
details are provided in \citet{allen11}; here we provide a summary of the method and the necessary modifications.

Before any analyses were performed on the data, all SDSS, Gemini and WHT spectra were corrected for the effects of Milky Way Galactic Extinction using the method of \citet{cardelli89} with A$_{V}$ values
taken from \citet{schlegel98}. All following processing steps were all performed in the quasar rest frame, using the redshifts derived by \citet{hewett10}.

Input samples of up to 500 spectra in each $\Delta z = 0.1$ redshift bin were drawn from the sample of SDSS DR6 quasar spectra \emph{without} broad absorption. Each of these input samples was processed using NMF to produce a set of 8--14 component spectra. These components can be linearly combined to reconstruct the spectrum of each of the input quasars. The number of components was selected independently for each redshift bin, and was set to be the greatest number that could be generated before overfitting occurred. Overfitted spectra are identified as those having much lower $\chi_{\nu}^{2}$ values than others in the input sample.

Once the component spectra had been generated, they could then be fitted to each BALQSO spectrum. Because the component spectra were based on non-BAL quasar spectra, the resulting reconstructions are of
the unabsorbed pseudo continuum. To perform this fit, the Gemini and WHT spectra were placed on the SDSS logarithmic wavelength scale, using a linear interpolation between pixels. An iterative procedure
was used to mask out the absorption regions from the fit. At each step in this procedure, the component spectra were fit to the observed spectrum using the current mask, and a new mask was generated by
searching for regions where the observed flux fell significantly below the reconstruction. The iteration continued until the change in the mask from one step to the next was small.

For the SDSS spectra, the most highly reddened quasars were given an empirical dust correction, assuming a power-law reddening curve, to match their overall shape to that of a typical quasar at the same
redshift. After the component spectra were fitted to the de-reddened spectrum, both the BALQSO spectrum and the reconstruction had the dust correction removed, matching the reconstruction to the shape of the original observed spectrum. For the Gemini and WHT spectra, the spectral coverage was insufficient to give reliable measurements of the continuum slope. However, the three objects in the original sample
whose SDSS continua required a reddening correction (J032832.77$-$070750.3, J034946.61$-$065730.3 and J111516.08$+$460234.6) were all excluded for other reasons; we hence assumed that no such corrections were required for the Gemini and WHT spectra of the remaining objects.

For the SDSS spectra, unphysical reconstructed emission line profiles in C\,{\sc iv} were detected automatically and corrected by reducing the number of components used in the fit. Additionally, spectra that failed certain automated quality control measurements were visually inspected for failures in the automatic masking procedure, and in such cases a manually-defined mask was applied and the components
re-fitted. For the Gemini and WHT data, the small number of spectra meant that they could all be visually inspected for both of these potential problems. The same corrections as for SDSS spectra were
applied in the cases where unphysical emission line profiles or poorly-defined masks were seen. Of the 35 quasars with variable BALs, three required corrections because of their emission line profiles and
nine because of poor automatic masks. The higher rate of manual intervention relative to SDSS spectra is primarily due to the smaller wavelength range of the Gemini spectra, and does not affect the
quality of the final reconstructions.

The uncertainty on the reconstructed flux level was determined using the synthetic BALQSO spectra described in Section 7.1 of \citet{allen11}. A sample of 50 non-BAL quasars was selected in each redshift bin, and a series of BAL troughs with known depth and shape were inserted into each spectrum. Extra noise was also added. The modified spectra were run through the same pseudo continuum fitting procedure as the observed spectra. The fits derived in this way were then compared to the fits produced for the original unabsorbed spectra in the wavelength range of the inserted BAL trough, defined as all pixels with a (known) flux ratio $< 0.9$. We characterised the error in each spectrum using the root mean square (RMS) of the fractional difference between reconstructions, then took the median value of this
RMS over all spectra in each bin of redshift and S/N. The resulting uncertainty was assumed to be dominated by the uncertainty in the normalisation of the pseudo continuum, rather than its shape, resulting in the errors in all pixels being perfectly correlated with each other.

\begin{figure}
\centering
\resizebox{\hsize}{!}{\includegraphics[angle=0,width=8cm]{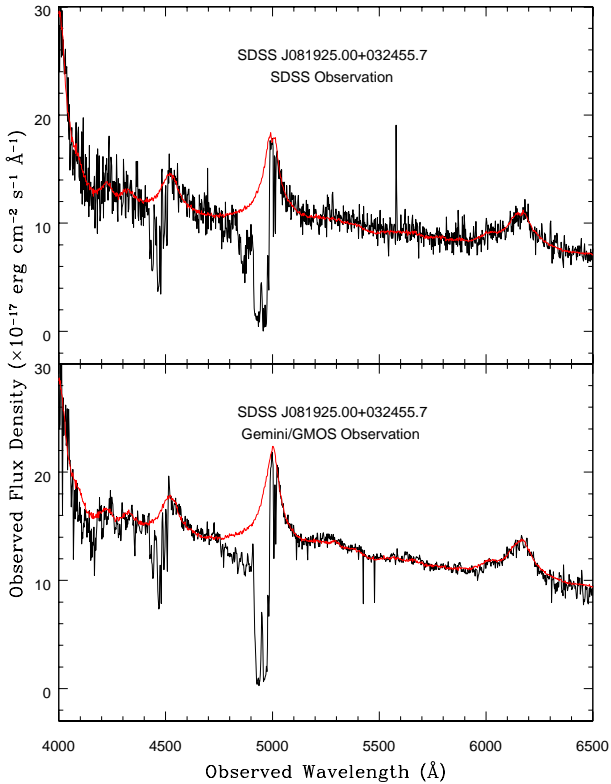}}
\caption{Top panel: Example SDSS BALQSO spectrum (black) and corresponding reconstruction (red). Bottom panel: Gemini/GMOS spectrum (black) and corresponding NMF reconstruction (red) of the same object. This quasar has redshift of z=2.138}
\label{fig:receg}
\end{figure}

The above steps produced a reconstruction of the unabsorbed pseudo continuum for each observed spectrum. For a given BALQSO, the reconstructions of the two epochs were derived in a consistent fashion, having used the same NMF components and the same fitting algorithms, but are independent of each other. Example reconstructions for the two observations of a single BALQSO are shown in Fig.~\ref{fig:receg}.

\section{Analysis}

	\subsection{BAL Identification}

Within our BALQSO sample, individual Si~{\sc iv} and C~{\sc iv} BALs were identified using a modified BI, differing from the original BI in that a lower velocity limit of 0 km s$^{-1}$ relative to the observed wavelength of the red doublet component of the appropriate ion emission line was used, together with an undefined upper velocity boundary (in practice no C~{\sc iv} absorption features are identified blueward of the Si~{\sc iv} emission line, corresponding to an outflow velocity of $\approx$-30000 km s$^{-1}$). A positive value of this modified BI for a particular absorption region categorised it as a BAL.

To carry out this procedure, the SDSS spectra and their Gemini or WHT counterparts were first resampled onto a 2.2\AA{} grid and smoothed using a 3 pixel boxcar smoothing algorithm to reduce the effects of
noise on the determination of BAL wavelength ranges. This smoothing was applied only for the purposes of BAL identification and continuum fitting, it was not used in subsequent investigations. A power law
continuum was fitted to both the SDSS and Gemini or WHT spectra using four spectral bands in each object judged to be relatively line free. Given the limited wavelength span of the Gemini observations, a
liberal approach was taken to this fitting, allowing wavelength bands in quasar rest-frame wavelength ranges of 1250-1350, 1600-1800, 1950-2050 and 2150-2250\AA{}, the only restriction being that at least
one of the four bands must be in the range 1250-1350\AA{} and at least one must be in any of the other three. As a result, the quasar SDSSJ032832.77-070750.3 was removed from our analysis as it contains
no identifiable continuum regions in the 1250-1350\AA{} range. The modeled continuum has the form F$_{\lambda}$=C$\lambda$$^{-\alpha}$ where C and $\alpha$ were calculated using an iteratively updating
procedure to minimise the chi-squared value of the fit. Three examples of BALQSO spectra and their modeled continua are shown in Fig. 2.

The SDSS spectrum and its modeled continuum for each object were examined by eye to search for regions of absorption which were plausible Si~{\sc iv} or C~{\sc iv} BAL candidates, the upper and lower boundaries of which correspond to the points in velocity space at which the spectrum crosses the modeled continuum. These regions were then tested using a BAL identification algorithm to determine
their status (BAL or non-BAL) based on the modified BI. From the total quasar sample, 38 Si~{\sc iv} and 59 C~{\sc iv} BALs were identified, these were then subjected to the variability testing stage (see
Section 4.2). A total of 38 objects showed lines from both transitions while 12 contained C~{\sc iv} BALs only (there were no quasars in the sample in which a Si~{\sc iv} BAL was unaccompanied by a C~{\sc iv} BAL). Some BALs were identified exclusively in the epoch 2 (Gemini or WHT) spectra, these were investigated in Section 5.3 as examples of appearing BALs.

\begin{figure}
\centering
\resizebox{\hsize}{!}{\includegraphics[angle=0,width=8cm]{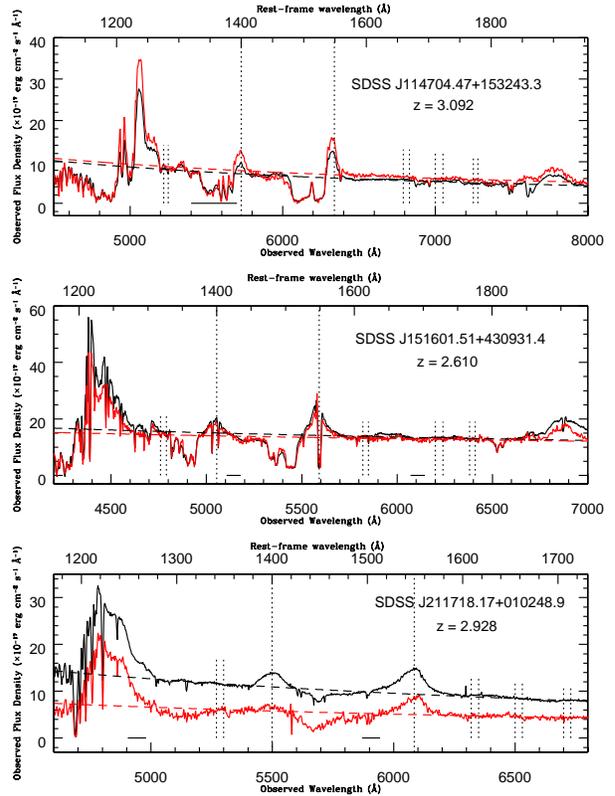}}
\caption{Power-law continuum fits to three example BALQSOs from our   sample. Shown are SDSS-observed epoch 1 (red) and epoch 2 (black) spectra. Continuum models (dashed lines) are fitted to spectral bands (bracketed by short vertical dotted lines). The Si~{\sc iv}  $\lambda$1400 and C~{\sc iv} $\lambda$1549 broad emission lines are indicated by long vertical dotted lines. The top panel object's epoch 2 observations were undertaken with WHT/ISIS, the middle and lower panel objects were observed in epoch 2 with Gemini/GMOS. Black horizontal lines indicate the spectral region of the combined ISIS dichroic overlap region (top panel) or the GMOS chip gaps (middle and lower panels).}
\label{fig:continuum}
\end{figure}

	\subsection{Variability Testing}

Prior to testing for BAL variability, each object had the higher resolution spectrum of the two epochs undergo smoothing by convolving with a Gaussian of appropriate full-width half maximum (FWHM) in order
to approximately degrade to the resolution of the object's other spectrum. The two spectra for each object were then resampled onto a common 2.2\AA{} grid for those observed with Gemini/GMOS, closely
matching the maximum linear wavelength bin width of the lower resolution SDSS spectra, and 3.7\AA{} for those taken with WHT/ISIS, which is approximately the width of the two-pixel binning in the WHT/ISIS spectra. These wavelength intervals correspond to velocity space intervals, for the redshift range covered in our sample, of between $\approx$85 and $\approx$140 km s$^{-1}$ for the Gemini spectra and $\approx$140 to $\approx$240 km s$^{-1}$ in the WHT/ISIS spectra (measured at a typical BAL absorber location of $\approx$-10000 km s$^{-1}$ from the C~{\sc iv} emission line). Small differences (a few \AA{}) in wavelength calibration between the SDSS and Gemini or WHT spectra were removed by aligning narrow emission or absorption features within each object in the Gemini or WHT spectra to those in the SDSS spectra where a consistent difference between several of these features was obvious.

The EW of each BAL was then calculated at both epochs based on the fitted continuum used for BAL identification, propagating errors during the procedure using the SDSS and Gemini or WHT error spectra and assuming a 10 $\%$ continuum error. A BAL was classed as variable if the significance of the EW difference between the two epochs exceeded 2.5$\sigma$. Adopting this approach, a total of 22 Si~{\sc iv} and 27 C~{\sc iv} BALs from 35 quasars were identified as variable, giving the percentage of varying BALs as 58$\%$ for Si~{\sc iv} and 46$\%$ for C~{\sc iv}. This subsample includes 14 objects which contain one variable BAL of each ion, allowing for comparisons of behaviour between both types of BAL within these particular objects.

	\subsection{Pseudo Equivalent Width}

In order to investigate the variability properties of the variable BALs, we define a modified equivalent width using Equation 2, which is referred to hereafter as a 'pseudo equivalent width'. This quantity is normalised to the NMF reconstructions (the pseudo continuum) as opposed to the modeled continuum EW used for BAL identification.

\begin{equation}
\label{eqn:pew}
EW_{p}=\int \frac{f_{cp}-f_{l}}{f_{cp}}dv
\end{equation}

\noindent In this case, $EW_{p}$ is the pseudo equivalent width, $f_{l}$ is the line flux density and $f_{cp}$ is the pseudo continuum flux density (in each velocity bin). The depth $f_{cp}-f_{l}$ at each
point in velocity space is therefore the difference between the absorption line flux and the sum of emission and continuum flux thought to be entering the absorber. Both transitions have doublet structures, with Si~{\sc iv} having a larger separation (1900 km s$^{-1}$) than C~{\sc iv} (500 km s$^{-1}$). This may cause Si~{\sc iv} BALs to be slightly wider than C~{\sc iv} BALs of the same intrinsic width but should not affect the validity of BAL variability measurements taken over the entire absorption line. We adopt conventions of measuring the velocities of blueshifted absorption features as negative, with pseudo equivalent widths being positive for regions of absorbed pseudo continuum. However when referring to the velocities of absorbers, terms such as greater or larger refer to the magnitude of the velocity, i.e. $|$$v$$|$.

We first translate observed wavelengths to quasar rest-frame velocities, with the laboratory wavelength of the red line of the doublet belonging to the appropriate transition defined as zero velocity. This ensures velocities of outflowing absorbing features will always be measured as negative at any velocity greater than zero. These zero velocities correspond to positions in wavelength space of 1402.77\AA{} for Si~{\sc iv} and 1550.77\AA{} for C~{\sc iv}. At each bin, wavelengths are converted to velocities using the relativistic Doppler shift formula

\begin{equation}
\label{eqn:doppler}
\frac{v}{c}=\frac{\left(\frac{\lambda_{line}}{\lambda_{abs}}\right)^{2}\left(1+z\right)^{2}-1}{\left(\frac{\lambda_{line}}{\lambda_{abs}}\right)^{2}\left(1+z\right)^{2}+1}
\end{equation}

\noindent where $\lambda_{line}$ is the rest-frame wavelength and $\lambda_{abs}$ is the observed wavelength of the absorption feature. Pseudo EW values are calculated by summing the products of pseudo continuum-normalised depths and velocity bin widths $\Delta{}v_{i}$ over the BAL velocity range $v_{min}$ to $v_{max}$ as follows

\begin{equation}
\label{eqn:pewcalc}
EW_{p}=\sum_{i=v_{min}}^{v_{max}}\left[\left(1-\frac{f(i)}{f_{cp}(i)}\right)\Delta{}v_{i}\right]
\end{equation}

\noindent By utilising the fractional error estimates in the position of the pseudo continuum described in Section 3 as well as the SDSS, Gemini/GMOS and WHT/ISIS error spectra, a total pseudo EW error is obtained. This consists of two functions $A$ and $B$ as follows

\begin{equation}
\label{eqn:pewerr1}
A=\sum_{i=v_{min}}^{v_{max}}\left[\left(\frac{\sigma_{f}(i)}{f_{cp}(i)}\right)^{2}\Delta{}v_{i}^{2}\right]
\end{equation}

\begin{equation}
\label{eqn:pewerr2}
B=\sum_{i=v_{min}}^{v_{max}}\left[\left(\frac{f(i)}{f_{cp}(i)}\right)\left(\frac{\sigma_{f_{cp}}(i)}{f_{cp}(i)}\right)\Delta{}v_{i}\right]
\end{equation}

\noindent The total error in the pseudo EW calculation is then

\begin{equation}
\label{eqn:pewerr}
\sigma_{EW_{p}}=\sqrt{A+B^{2}}
\end{equation}

	\subsection{BAL pseudo EW Variability}

Following L07 (for pseudo EW rather than continuum-based EW), we measured the fractional change in BAL pseudo EW, $\Delta$$EW_{p}$$/$$\langle$$EW_{p}$$\rangle$, where $\Delta$$EW_{p}$ is the difference in pseudo EW and $\langle$$EW_{p}$$\rangle$ is the average pseudo EW across the two epochs. Fractional pseudo EW changes are a robust method of measuring variability since they represent, for a mean value across the two epochs, a change in the BAL gas covering fraction or the fractional change, driven by ionizing continuum fluctuations, in the absorbing ion population for a non-saturated absorber. Comparisons of fractional pseudo EW variations for an object which contains BALs from both ions and which overlap in velocity space can provide insight into outflow structure and possible contributions
of ionization changes or covering fraction variations. A summary of all variable BALs, including velocity boundaries ($v_{max}$ and v$_{min}$), the pseudo EWs at each epoch, the time interval between epochs in the quasar rest frame ($t_{qrest}$) and the fractional pseudo EW change $\Delta$$EW_{p}$$/$$\langle$$EW$$\rangle$ is presented in Table 2.

\begin{table*}
\begin{center}
\caption{Two-epoch pseudo equivalent widths and fractional variability of variable BALs.}
\begin{tabular}{l l l l l l l l}
\hline Object (BAL No.)&$\Delta$$t_{qrest}$&Ion&$v_{max}$&$v_{min}$&$EW_{p1}$&$EW_{p2}$&$\Delta$$EW_{p}$$/$$\langle$$EW_{p}$$\rangle$\\
 &(days)& &(km $s^{-1}$)&(km s$^{-1}$)&(km s$^{-1}$)&(km s$^{-1}$)& \\
\hline
SDSSJ001025.90+005447.6&1052&C~{\sc iv}&-13200&-8200&2941$\pm$240&3361$\pm$77&0.133$\pm$0.080\\
SDSSJ004613.54+010425.7&1282&Si~{\sc iv}&-20400&-3000&2986$\pm$380&2245$\pm$243&-0.284$\pm$0.174\\
 & &C~{\sc iv}&-28100&-6300&5449$\pm$575&4514$\pm$276&-0.188$\pm$0.129\\
SDSSJ025720.43-080322.5&1332&Si~{\sc iv}&-25500&-21800&1222$\pm$502&1121$\pm$135&-0.086$\pm$0.444\\
SDSSJ031033.45-060957.8&1296&Si~{\sc iv}&-20200&-14800&673$\pm$319&1253$\pm$76&0.602$\pm$0.355\\
SDSSJ031331.22-070422.8&1064&Si~{\sc iv}&-17100&-5600&5213$\pm$178&5638$\pm$53&0.078$\pm$0.034\\
 & &C~{\sc iv}&-17400&-5800&7634$\pm$199&8292$\pm$47&0.083$\pm$0.026\\
SDSSJ033223.51-065450.5&830&Si~{\sc iv}&-19500&-10100&3595$\pm$384&3343$\pm$75&-0.073$\pm$0.113\\
SDSSJ035749.11-061121.9&1328&C~{\sc iv}&-13600&-7000&2394$\pm$178&656$\pm$102&-1.139$\pm$0.155\\
SDSSJ073535.44+374450.4&994&Si~{\sc iv}&-19800&-5000&3729$\pm$238&5074$\pm$87&0.305$\pm$0.058\\
 & &C~{\sc iv}&-28000&-7500&7716$\pm$355&8978$\pm$99&0.151$\pm$0.044\\
SDSSJ081823.46+484910.8&1252&C~{\sc iv}&-14100&-5300&3703$\pm$488&2512$\pm$411&-0.383$\pm$0.209\\
SDSSJ081925.00+032455.7&1032&Si~{\sc iv}&-9000&-4000&1686$\pm$148&1109$\pm$44&-0.412$\pm$0.113\\
SDSSJ083718.63+482806.1&804&Si~{\sc iv}&-16100&-4500&5210$\pm$177&6132$\pm$92&0.163$\pm$0.035\\
SDSSJ084408.29+423226.9&775&Si~{\sc iv}&-17800&-5200&2825$\pm$145&3364$\pm$80&0.174$\pm$0.054\\
 & &C~{\sc iv}&-21300&-4700&5712$\pm$436&6475$\pm$86&0.125$\pm$0.073\\
SDSSJ085006.08+072959.0&819&C~{\sc iv}&-27300&-13900&1313$\pm$188&0$\pm$113&-2.000$\pm$0.472\\
SDSSJ085104.05+051539.8&446&Si~{\sc iv}&-12600&-5300&2112$\pm$176&3662$\pm$32&0.537$\pm$0.064\\
 & &C~{\sc iv}&-11300&-4000&5578$\pm$87&5814$\pm$21&0.041$\pm$0.016\\
SDSSJ092639.34+383656.7&912&C~{\sc iv}&-22800&-8600&3772$\pm$645&2841$\pm$185&-0.282$\pm$0.205\\
SDSSJ093251.98+023727.0&1272&Si~{\sc iv}&-30700&-6300&1086$\pm$901&2536$\pm$648&0.801$\pm$0.660\\
 & &C~{\sc iv}&-29400&-11000&6211$\pm$475&7443$\pm$179&0.180$\pm$0.075\\
SDSSJ100312.63+402505.6&606&Si~{\sc iv}&-13400&-7200&2503$\pm$277&2857$\pm$72&0.132$\pm$0.107\\
SDSSJ101056.68+355833.3&687&C~{\sc iv}&-13100&-3400&6407$\pm$149&6333$\pm$127&-0.012$\pm$0.031\\
SDSSJ104059.79+055524.4&963&Si~{\sc iv}&-15600&-4500&2295$\pm$396&4202$\pm$55&0.587$\pm$0.128\\
SDSSJ110041.19+003631.9&1346&Si~{\sc iv}&-17000&-5500&2909$\pm$435&633$\pm$185&-1.285$\pm$0.317\\
 & &C~{\sc iv}&-22100&-5000&7029$\pm$505&3246$\pm$233&-0.736$\pm$0.115\\
SDSSJ110339.90+011928.5&1173&Si~{\sc iv}&-10800&-4800&1909$\pm$400&2871$\pm$128&0.403$\pm$0.179\\
SDSSJ112733.69+343008.8&436&C~{\sc iv}&-19500&-11100&2377$\pm$418&2857$\pm$129&0.184$\pm$0.168\\
SDSSJ113831.42+351725.3$^{\dagger}$&360&Si~{\sc iv}&-14100&-1600&4381$\pm$621&1501$\pm$180&-0.979$\pm$0.245\\
 & &C~{\sc iv}&-20900&-1700&8905$\pm$782&4764$\pm$236&-0.606$\pm$0.125\\
SDSSJ114704.47+153243.3$^{\dagger}$&303&Si~{\sc iv}&-20600&-3200&7354$\pm$403&7157$\pm$145&-0.027$\pm$0.059\\
 & &C~{\sc iv}&-27300&-3200&10323$\pm$620&7918$\pm$200&-0.264$\pm$0.072\\
SDSSJ134458.82+483457.5$^{\dagger}$&615&C~{\sc iv}&-21100&-7100&2688$\pm$1058&5493$\pm$319&0.686$\pm$0.288\\
SDSSJ142244.45+382330.6&311&C~{\sc iv}&-16500&-7400&2542$\pm$239&3808$\pm$60&0.399$\pm$0.079\\
SDSSJ143604.64+350428.5&616&Si~{\sc iv}&-20600&-10800&4342$\pm$663&3837$\pm$238&-0.124$\pm$0.172\\
 & &C~{\sc iv}&-16800&-7000&7828$\pm$557&5146$\pm$183&-0.413$\pm$0.092\\
SDSSJ151601.51+430931.4&691&Si~{\sc iv}&-15300&-5700&3633$\pm$196&4169$\pm$85&0.138$\pm$0.055\\
 & &C~{\sc iv}&-25800&-4600&5712$\pm$380&6755$\pm$122&0.167$\pm$0.064\\
SDSSJ162657.47+405848.0$^{\dagger}$&551&C~{\sc iv}&-13100&-2500&5884$\pm$238&4886$\pm$107&-0.185$\pm$0.049\\
SDSSJ164152.30+305851.7$^{\dagger}$&611&Si~{\sc iv}&-30400&-1800&26259$\pm$1468$^{*}$&26519$\pm$212$^{*}$&0.010$\pm$0.056$^{*}$\\
 & &C~{\sc iv}&-27600&-1800&13166$\pm$1166&13886$\pm$212&0.053$\pm$0.088\\
SDSSJ210436.62-070738.3&1080&C~{\sc iv}&-23500&-9800&2674$\pm$462&5922$\pm$58&0.756$\pm$0.116\\
SDSSJ211718.17+010248.9&860&Si~{\sc iv}&-27700&-13100&1931$\pm$512&292$\pm$121&-1.475$\pm$0.588\\
 & &C~{\sc iv}&-25800&-15100&3348$\pm$372&664$\pm$85&-1.338$\pm$0.229\\
SDSSJ212412.60+095923.3$^{\dagger}$&737&Si~{\sc iv}&-13300&-3300&1054$\pm$966&227$\pm$450&-1.291$\pm$1.981\\
 & &C~{\sc iv}&-14000&-2900&6299$\pm$515&3872$\pm$434&-0.477$\pm$0.136\\
SDSSJ213138.93-070013.3&1188&C~{\sc iv}&-27400&-3800&5368$\pm$1185&5697$\pm$557&0.059$\pm$0.237\\
SDSSJ222505.28-084542.7&1176&C~{\sc iv}&-26900&-7500&4464$\pm$1165&2380$\pm$467&-0.609$\pm$0.384\\
\hline
\end{tabular}
\label{tab:bals}
\end{center}
$^{\dagger}$ These objects were observed using WHT/ISIS, unmarked objects were observed using Gemni/GMOS.\\
$^{*}$ May include contamination from C~{\sc ii} LoBAL.\\
Outflow velocities are rounded to the nearest 100 km s$^{-1}$ as the error in velocity width is $\approx$200 km s$^{-1}$.\\
$EW_{p1}$ and $EW_{p2}$ denote pseudo EW for earlier and later epochs respectively.
\end{table*}

The timescales in this study represent an intermediate epoch rest frame separation time between the C~{\sc iv} short term data in L07 of a few weeks to months and the mostly longer timescales investigated in \citet{gibson08} (hereafter referred to as G08) and the timescales between the Large Bright Quasar Survey (LBQS) and Hobby-Eberly Telescope (HET) observations in \citet{gibson10} (hereafter referred to as G10). The G08 study produced results based on a 3 to 6 year quasar rest frame epoch separation time for C~{\sc iv} only. G10, on the other hand, was a multi-epoch study with the longest rest frame time period being $\approx$5 to 7 years additionally including the Si~{\sc iv} absorption region. Notably, G10 only had full spectral coverage (defined as 0 to 30000 km s$^{-1}$) for Si~{\sc iv} in the LBQS and HET spectra and so could only provide results for the Si~{\sc iv} absorption region on corresponding time scales.

Fig. 3 illustrates the fractional change in EW of BALs in our sample plotted against rest frame time interval. A detailed investigation of the trends in variability with respect to time and other variables is reported in the discussion. 

\begin{figure}
\centering
\resizebox{\hsize}{!}{\includegraphics[angle=0,width=8cm]{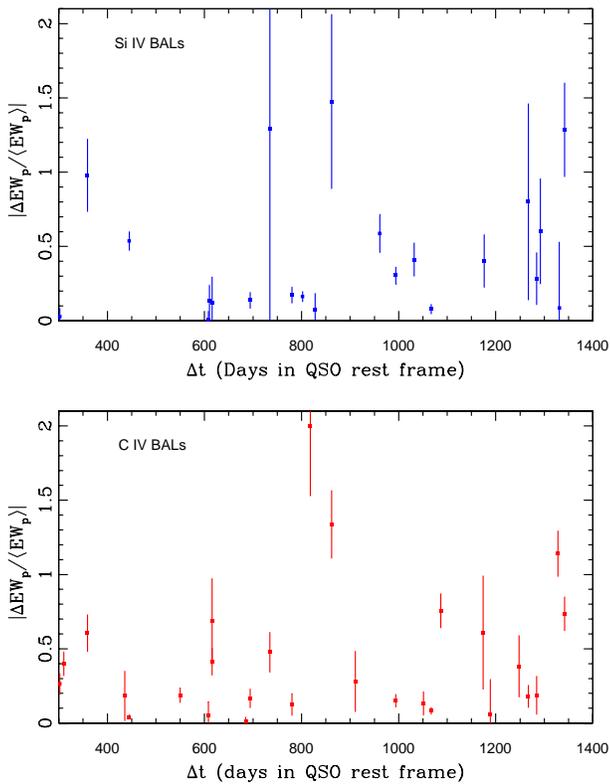}}
\caption{Magnitude of fractional change in pseudo EW against quasar rest frame time interval between epochs for Si~{\sc iv} (upper panel) and C~{\sc iv} (lower panel).}
\label{fig:ewtime}
\end{figure}

\section{Discussion}

	\subsection{Trends, comparisons of C~{\sc iv} and Si~{\sc iv} properties and variability}

Attempts to understand the physical mechanisms giving rise to BAL variability require an investigation into the correlations, if any, of the amplitude of variation with other properties, including the luminosity of the host quasar, elapsed rest frame time between observations, the velocity of the bulk of the outflowing gas and BAL pseudo EW (comprising depth, width and velocity). To achieve this, a series of Spearman rank correlation tests were carried out to determine the likelihood of dependence of $|$$\Delta$$EW_{p}$$/$$\langle$$EW_{p}$$\rangle$$|$ and $|$$\Delta$$EW_{p}$$|$ on such parameters, the results of which are presented in Table 3. This was achieved by calculating the p-value for the likelihood of a null hypothesis (no underlying correlation) and subtracting it from 100$\%$ probability. A correlation probability is considered highly significant if it is greater than 99$\%$. Significance values of between 95$\%$ and 99$\%$ are considered plausible but no definitive conclusions are drawn from such values.

Splitting the variability dependence tests into correlations of fractional change and absolute change in $EW_{p}$ is useful as the former could indicate that a parameter's increase or decline correlates with the fraction of outflowing gas along the line of sight affected by the variability inducing effect (possibly ionization fraction changes or changes in covering fraction). The latter may indicate that the parameter change is correlated with absolute changes in line of sight column density. 

\begin{table*}
\begin{center}
\caption{Results of Spearman rank Correlation Tests}
\begin{tabular}{l l l}
\hline Comparison&Si~{\sc iv} Confidence (Sign of $\rho$)&C~{\sc iv} Confidence (Sign of $\rho$)\\
\hline
$|$$\Delta$$EW_{p}$$|$ vs. $\langle$$EW_{p}$$\rangle$&72$\%$$^{b}$ (-)&41$\%$ (-)\\
$|$$\Delta$$EW_{p}$$/$$\langle$$EW_{p}$$\rangle$$|$ vs. B$_{vel}$&78$\%$ (-)&93$\%$ (+)\\
$|$$\Delta$$EW_{p}$$|$ vs. B$_{vel}$&92$\%$ (-)&71$\%$ (+)\\
$|$$\Delta$$EW_{p}$$/$$\langle$$EW_{p}$$\rangle$$|$ vs. $\langle$B$_{depth}$$\rangle$&$>$99.99$\%$$^{b}$ (-)&99.95$\%$ (-)\\
$|$$\Delta$$EW_{p}$$|$ vs. $\langle$B$_{depth}$$\rangle$&98$\%$$^{b}$ (-)&93$\%$ (-)\\
$|$$\Delta$$EW_{p}$$/$$\langle$$EW_{p}$$\rangle$$|$ vs. M$_{i}$&61$\%$ (+)&93$\%$ (+)\\
$|$$\Delta$$EW_{p}$$|$ vs. M$_{i}$&6$\%$ (-)&97$\%$ (+)\\
$|$$\Delta$$EW_{p}$$/$$\langle$$EW_{p}$$\rangle$$|$ vs. t$_{qrest}$&78$\%$ (+)&63$\%$ (+)\\
$|$$\Delta$$EW_{p}$$|$ vs. t$_{qrest}$&51$\%$ (+)&15$\%$ (+)\\
$|$$\Delta$$EW_{p}$$/$$\langle$$EW_{p}$$\rangle$$|$ vs. V$_{width}$&40$\%$$^{b}$ (+)&40$\%$ (-)\\
$|$$\Delta$$EW_{p}$$|$ vs. V$_{width}$&83$\%$$^{b}$ (+)&58$\%$ (+)\\
\hline    
\end{tabular}
\label{tab:sptest}
\end{center}
$^{b}$The unusual Si~{\sc iv} $\langle$$EW_{p}$$\rangle$ measurement for J164152.30+305851.7 is not included in this test.\\
\end{table*}

The fractional change in pseudo EW with respect to quasar rest frame time scale does not show a significant correlation, with only 63$\%$ confidence for C~{\sc iv} and 78$\%$ for Si~{\sc iv}. This indicates
that time dependence of variability amplitude is not visible in our sample over the timescales investigated. The absolute change $\Delta$$EW_{p}$ is also not correlated with rest frame time interval in either ion. It is worth considering whether or not the rest frame timescales have an impact on the variability categorisation of the BALs. The 35 quasars containing BALs classed as variable have a mean
t$_{qrest}$ of 757 days, compared with 833 days between observations for the 15 quasars containing no variable BALs. This difference is not considered to be significant as the standard deviation on the range of rest frame timescales of variable quasars is $\sigma_{\Delta t}$=316 days. These findings contrast with studies on shorter and longer rest-frame timescales \citep{lundgren07,gibson10} in which the most variable absorption features were concentrated at the longest rest-frame time intervals.

Quasars containing variable BALs have a mean luminosity of -27.583 M$_{i}$ compared to a mean of -27.307 M$_{i}$ for quasars containing no variable BALs. As in the case of time separation, this is not
considered to be a significant difference since, for the distribution of luminosities in variable quasars, $\sigma_{Lum}$=0.626 M$_{i}$.

SDSSJ164152.30+305851.7 is the only object in our sample which shows evidence for LoBAL absorption features. Obvious examples of these features are visible in both the WHT and SDSS observations of this object and correspond to the transitions of Al~{\sc iii} at 1860\AA{} and Mg~{\sc ii} at 2798\AA{}. In addition to these, in \citet{hall02} a rarer LoBAL resulting from the C~{\sc ii} transition at 1334\AA{} was found in several LoBAL QSOs. A measurement of the average pseudo EW between -1800 and -30400 km s$^{-1}$ from the Si~{\sc iv} emission line centre gives an exceptionally large value of $\langle$$EW_{p}$$\rangle$=26389$\pm$742 km s$^{-1}$, several times that of the next strongest BAL in the Si~{\sc iv} sample. It therefore seems very plausible that this measurement is contaminated by additional C~{\sc ii} LoBAL absorption in the Si~{\sc iv} BAL absorption range, given that the centre of the C~{\sc ii} emission line is located approximately -15000 km s$^{-1}$ from the centre of the Si~{\sc iv} emission line. Further evidence for the possible existence of such absorption is provided by the fact that the Al~{\sc iii} LoBAL spans a velocity range of approximately 0 to -14000 km s$^{-1}$, providing the likely position of any C~{\sc ii} absorption (indicated in Fig. 4). For these reasons, the Si~{\sc iv} BAL from J164152.30+305851.7 is not included in the tests for dependence on average pseudo EW, or its component quantities of depth and width.

\begin{figure}
\centering
\resizebox{\hsize}{!}{\includegraphics[angle=0,width=8cm]{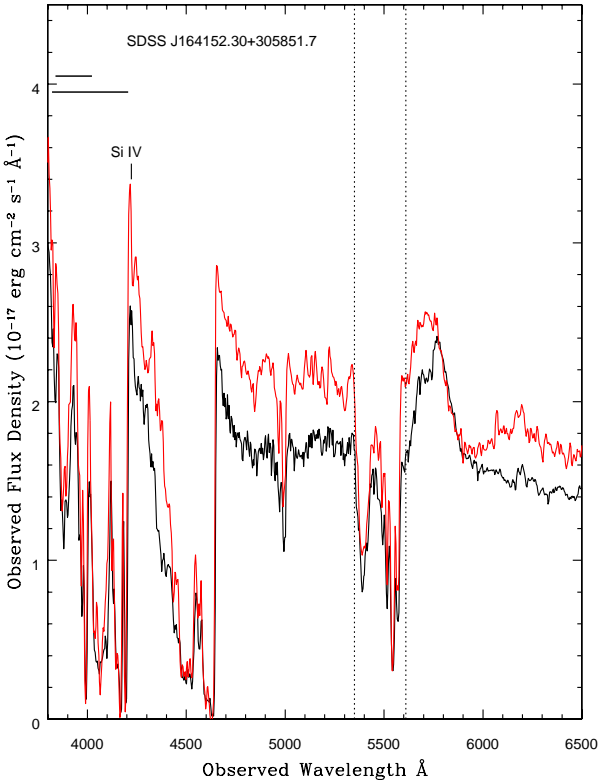}}
\caption{WHT/ISIS spectrum (black) and Gemini/GMOS spectrum (red) of the LoBAL quasar SDSS J164152.30+305851.7. Dotted vertical lines indicate the extent of Al~{\sc iii} LoBAL absorption (from zero to approximately -14000 km s$^{-1}$). The Si~{\sc iv} emission line is also labelled. The lower thick horizontal black line indicates the range of BAL absorption measured blueward of the Si~{\sc iv} emission line, while the upper thick horizontal black line indicates the possible location of C~{\sc ii} LoBAL absorption based on the extent of the Al~{\sc iii} absorption.}
\label{fig:fig164152}
\end{figure}

A significant correlation is not observed in the $|$$\Delta$$EW_{p}$$|$ vs. $\langle$$EW_{p}$$\rangle$ relationship, which the Spearman rank test gives as 72$\%$ confidence for Si~{\sc iv} and 41$\%$ for C~{\sc iv}. The value of $\langle$$EW_{p}$$\rangle$ can be split into two more fundamental properties, one being the velocity width (V$_{width}$) of the BAL, corresponding to the difference in velocity between the upper and lower boundaries of the BAL trough, the other being the depth ($\langle$B$_{depth}$$\rangle$), defined as the mean pseudo continuum normalised depth across the BAL trough averaged between the two epochs. Our test of $|$$\Delta$$EW_{p}$$/$$\langle$$EW_{p}$$\rangle$$|$ vs. V$_{width}$ does not give a significant correlation for either Si~{\sc iv} or C~{\sc iv} (significance of 40$\%$ for both ions). This is in stark contrast with the strong inverse correlation found for the $|$$\Delta$$EW_{p}$$/$$\langle$$EW_{p}$$\rangle$$|$ vs. $\langle$B$_{depth}$$\rangle$ dependence, which the Spearman rank test reveals to be $>$99.99$\%$ for the Si~{\sc iv} sample and 99.95$\%$ for the C~{\sc iv} sample, with the negative sign of $\rho$ indicating an inverse scaling (see Fig. 5). However this result is likely to be due to the fact that $\langle$$EW_{p}$$\rangle$ shows a strong positive correlation with $\langle$B$_{depth}$$\rangle$ in both Si~{\sc iv} (99.97$\%$ significance) and C~{\sc iv} (99.93$\%$ significance). Since the fractional pseudo EW measure incorporates $\langle$$EW_{p}$$\rangle$ as the denominator, such a correlation is not inferred to represent anything physical. Correlation probability is possibly significant (98$\%$) in the $|$$\Delta$$EW_{p}$$|$ vs. B$_{depth}$ relationship for Si~{\sc iv} BALs but not so in the C~{\sc iv} sample (93$\%$), with negative $\rho$ in both cases. This is similar to the findings of \citet{gibson08} for C {\sc iv} BALs, which found no correlation between the absolute change in absorption strength in velocity ranges which varied and the depth of those regions. There is no discernible correlation between $|$$\Delta$$EW_{p}$$|$ and V$_{width}$ (83$\%$ correlation confidence for Si~{\sc iv} and 58$\%$ for C~{\sc iv}). Previous work comparing variability to velocity width is minimal, although \citet{filizak12} found tentative evidence that relatively narrow C~{\sc iv} BALs are more likely to undergo disappearance than wider BALs. 

\begin{figure}
\centering
\resizebox{\hsize}{!}{\includegraphics[angle=0,width=8cm]{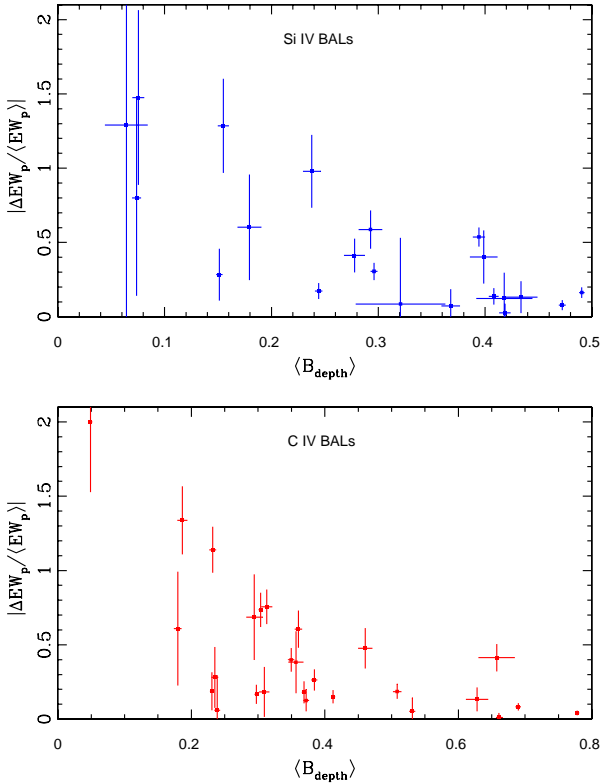}}
\caption{Fractional pseudo EW variability for Si~{\sc iv} BALs (top panel) and C~{\sc iv} BALs (bottom panel) vs. average mean normalised BAL depth across the two epochs. Outlying Si~{\sc iv} BAL measurement from J164152.30+305851.7 is not included.}
\label{fig:ewdepth}
\end{figure}

For this study, the BAL velocity (B$_{vel}$) is defined as the velocity of the depth centroid of the BAL feature, giving the best estimate of the value for the column density weighted mean of the velocity of the outflowing gas. A study of the dependence of variability amplitude on velocity across a range of velocity bins in a sample of C~{\sc iv} BALs in \citet{capellupo11} showed a relationship of strengthening variability with increased velocity. The test for the correlation of $|$$\Delta$$EW_{p}$$/$$\langle$$EW_{p}$$\rangle$$|$ on B$_{vel}$ in our dataset returns a probability of 78$\%$ for Si~{\sc iv} and 93$\%$ for C~{\sc iv}, implying that there is no discernible correlation. Although this is not exactly the same type of study as in \citet{capellupo11} since our work reports on statistics for the whole BAL features rather than portions thereof, this may suggest a contradiction with those previously reported results. For the $|$$\Delta$$EW_{p}$$|$ vs. B$_{vel}$ relationship, correlation significances of 92$\%$ for Si~{\sc iv} BALs and 71$\%$ for C~{\sc iv} BALs are found. 

The dependence or otherwise of the host quasar luminosity on BAL variability can shed light on the contribution of ionization fraction changes to the variability, as it is a general property of quasars
that emission line variability, which is known to be driven by ionization continuum variations, has an inverse dependence on host luminosity \citep{vandenberk04}. Luminosity is defined in terms of the
SDSS i-band absolute magnitude ($M_{i}$) as listed in \citet{shen11}. There is no strong evidence in this sample, in either C~{\sc iv} or Si~{\sc iv} BALs, of a correlation in the $|$$\Delta$$EW_{p}$$/$$\langle$$EW_{p}$$\rangle$$|$ vs. M$_{i}$ relationship, given correlation probabilities of 93$\%$ and 61$\%$ respectively. The Spearman rank test also gives a confidence of 6$\%$ for Si{\sc iv} and 97$\%$ for C~{\sc iv} correlation with respect to $|$$\Delta$$EW_{p}$$|$ vs. M$_{i}$, which, given a positive value of $\rho$, may indicate a inverse scaling between the absolute variability and luminosity for C~{\sc iv} BALs. BAL variability dependence on host quasar luminosity has not been extensively studied, however \citet{filizak12} reported no correlation between quasar luminosity and the rate of C~{\sc iv} BAL disappearance. 

\subsubsection{Quasars containing both Si~{\sc iv} and C~{\sc iv} variable BALs}

\begin{figure*}
\centering
\includegraphics[width=170mm]{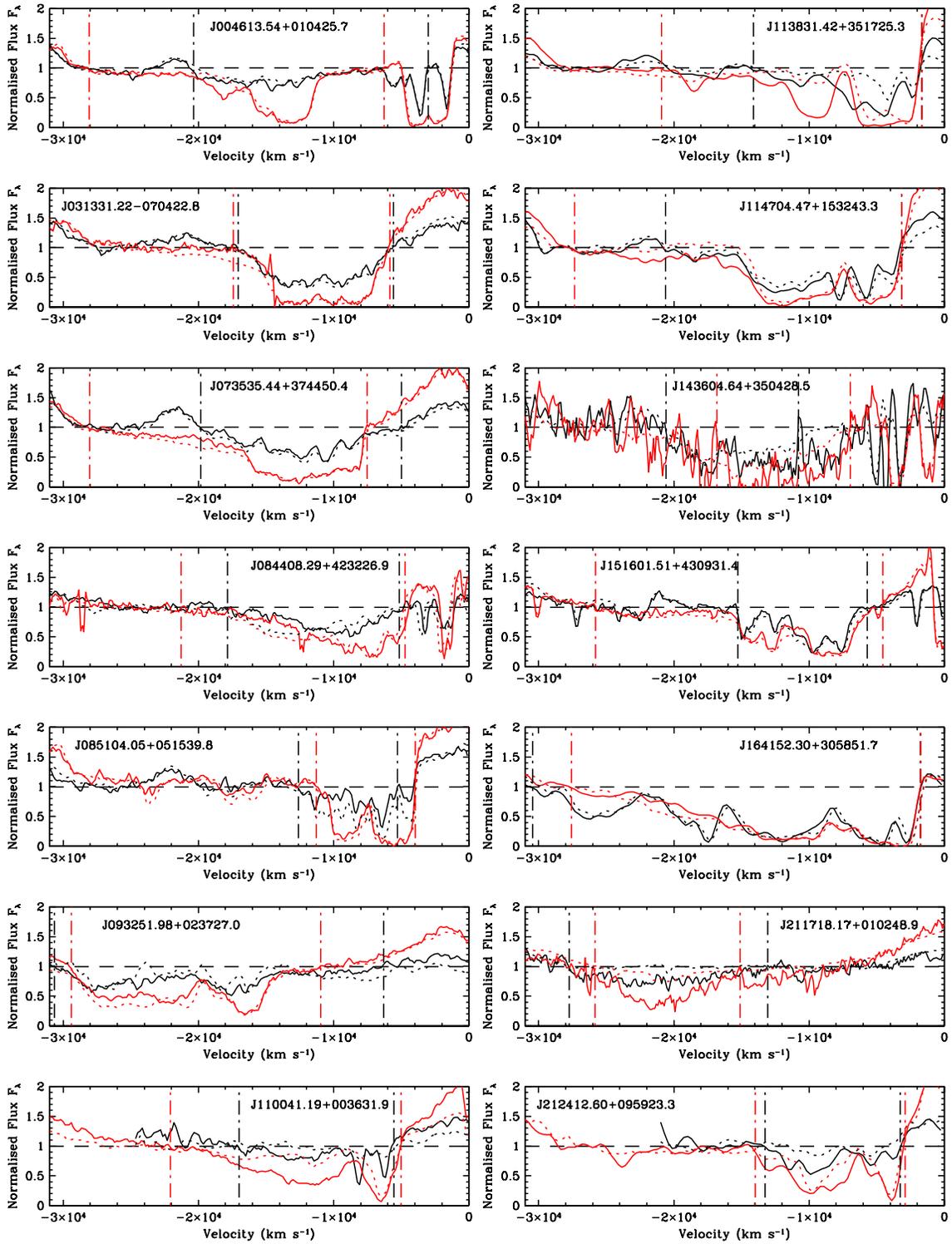}
\caption{Behaviour of Si~{\sc iv} (black) and C~{\sc iv} (red) BAL regions plotted in concurrent velocity space for each of the dual ion variable BAL quasar sample. Spectra are smoothed using a 3-pixel  boxcar smoothing for clarity. Solid lines represent the SDSS spectra while dotted lines represent the Gemini/WHT spectra. Vertical dash-dot lines represent the BAL boundaries. A can be seen, the overlap region forms the majority of the width of both BALs in every case.}
\label{fig:velcomp}
\end{figure*}

An interesting subsample of quasars containing variable BALs are those which contain examples in both Si~{\sc iv} and C~{\sc iv}, allowing direct comparison of the behaviour of these absorption features within the same objects. There are 14 such quasars in our sample, each consisting of one BAL from each ion. Two notable facts are apparent in this subsample, namely (1) for any given quasar, both BALs change in the same direction (as has been noted previously in \citet{capellupo12}) and (2) when considering the velocity range over which the BALs span, there is a considerable overlap between the C~{\sc iv} and Si~{\sc iv} outflows. This suggests that the same outflowing structure is responsible for such BALs and that the dominant variability mechanism is the same in each case. 

Comparisons of the average values of several BAL parameters between each ion can provide an insight into the likely physical mechanisms leading to variability. For example, if the outflows of one ion were
found to be at a significantly higher mean velocity than that of the other ion then, following the model of \citet{elvis00}, this would provide evidence for this ion having a greater density than the other
at positions closer to the origin of the outflow. Table 4 provides mean values of several properties in the dual ion variable BAL subsample, while Fig. 6 displays the spectra of these objects.

\begin{table}
\begin{center}
\caption{Mean BAL Properties for quasars containing variable BALs from both ions.}
\begin{tabular}{l l l}
\hline Property&Si~{\sc iv} mean&C~{\sc iv} mean\\
\hline
B$_{vel}$&-13181$\pm$55 km s$^{-1}$&-11762$\pm$53 km s$^{-1}$\\
B$_{vel}$*&-12271$\pm$56 km s$^{-1}$&-11880$\pm$54 km s$^{-1}$\\
$\langle$$EW_{p}$$\rangle$&4881$\pm$92 km s$^{-1}$&6738$\pm$80 km s$^{-1}$\\
$\langle$$EW_{p}$$\rangle$*&3226$\pm$80 km s$^{-1}$&6216$\pm$72 km s$^{-1}$\\
$|$$\Delta$$EW_{p}$$/$$\langle$$EW_{p}$$\rangle$$|$&0.536$\pm$0.159&0.345$\pm$0.028\\
$|$$\Delta$$EW_{p}$$/$$\langle$$EW_{p}$$\rangle$$|$*&0.577$\pm$0.171&0.367$\pm$0.030\\
V$_{width}$&14439$\pm$111 km s$^{-1}$&16812$\pm$107 km s$^{-1}$\\
V$_{width}$*&13345$\pm$113 km s$^{-1}$&16120$\pm$109 km s$^{-1}$\\
$\langle$B$_{depth}$$\rangle$&0.308$\pm$0.003&0.431$\pm$0.002\\
$\langle$B$_{depth}$$\rangle$*&0.262$\pm$0.003&0.423$\pm$0.003*\\
\hline    
\end{tabular}
\label{tab:avprops}
\end{center}
*SDSSJ164152.30+305851.7 excluded 
\end{table}

This sample includes the LoBAL quasar SDSSJ164152.30+305851.7 which contains a Si~{\sc iv} BAL region of exceptional large average pseudo EW of $\langle$$EW_{p}$$\rangle$=26389$\pm$742 km s$^{-1}$, which is not matched by a similarly large C~{\sc iv} BAL and may be due to C~{\sc ii} absorption contaminating the measurement. Table 4 therefore includes a duplicate set of comparisons from which SDSSJ164152.30+305851.7 is excluded. If all 14 quasars are taken into account, the mean BAL velocity of Si~{\sc iv} is larger than that of C~{\sc iv} by $\approx$1419 km~s$^{-1}$, however this reduces to a difference of $<$400 km s$^{-1}$, about twice the typical BAL velocity error, in the 13 quasar sample. This difference is therefore considered to be consistent with zero. The mean epoch averaged pseudo EW is larger in C~{\sc iv} than in Si~{\sc iv} by 38$\%$, which increases dramatically to 93$\%$ in the 13 quasar sample where the apparent large Si~{\sc iv} BAL from SDSSJ164152.30+305851.7 is excluded. This difference is mainly explained by the fact that C~{\sc iv} BALs are on average 40$\%$ deeper (61$\%$ in the 13 quasar sample) than Si~{\sc iv}. However, there is also a difference in velocity width, with the mean value for C~{\sc iv} being $\approx$2400 km s$^{-1}$ larger, or $\approx$2800 km s$^{-1}$ larger without the LoBAL quasar, than that for Si~{\sc iv}. Also, the Si~{\sc iv} BALs show a 55$\%$ (57 $\%$ for the  13 quasar sample) stronger mean fractional pseudo EW variability than C~{\sc iv}.

To determine what relationship, if any, exists between the variability of all the Si~{\sc iv} and C~{\sc iv} BALs in this quasar subsample, a Spearman Rank test was carried out to determine the significance of a correlation between the Si~{\sc iv} and C~{\sc iv} values of $|$$\Delta$$EW_{p}$$/$$\langle$$EW_{p}$$\rangle$$|$. This reveals a probability of 98$\%$ of correlation, just below the threshold of 99$\%$ for strong significance. The test was also performed on the dependence of $|$$\Delta$$EW_{p}$$|$ of Si~{\sc iv} to that of C~{\sc iv}, revealing a probability of correlation of 91$\%$, too low to be considered significant. This relationship between the fractional changes in BAL pseudo EW of the two ions is illustrated in Fig. 7 and shows that a least squares fit line closely matches 8 of the data points, including that of the LoBAL SDSSJ164152.30+305851.7 (marked "1641" in the plot). No properties of the BALs nor their host quasars were found that distinguished those
which fell relatively close to the best fit line from those which did not.

\begin{figure}
\centering
\resizebox{\hsize}{!}{\includegraphics[angle=0,width=8cm]{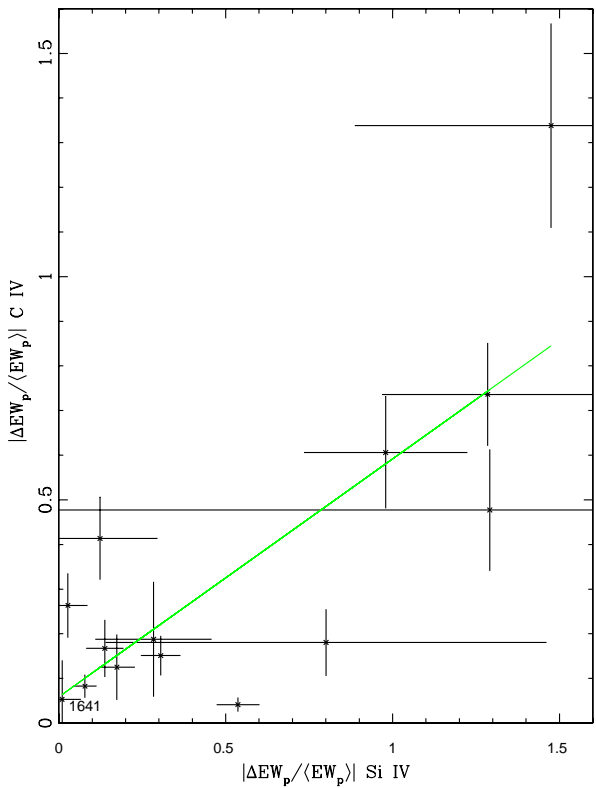}}
\caption{Fractional change in pseudo EW for C~{\sc iv} (vertical axis) and Si~{\sc iv} (horizontal axis). The green line represents a least squares fit to the data.}
\label{fig:varcomp}
\end{figure}

\begin{figure*}
\centering
\includegraphics[width=170mm]{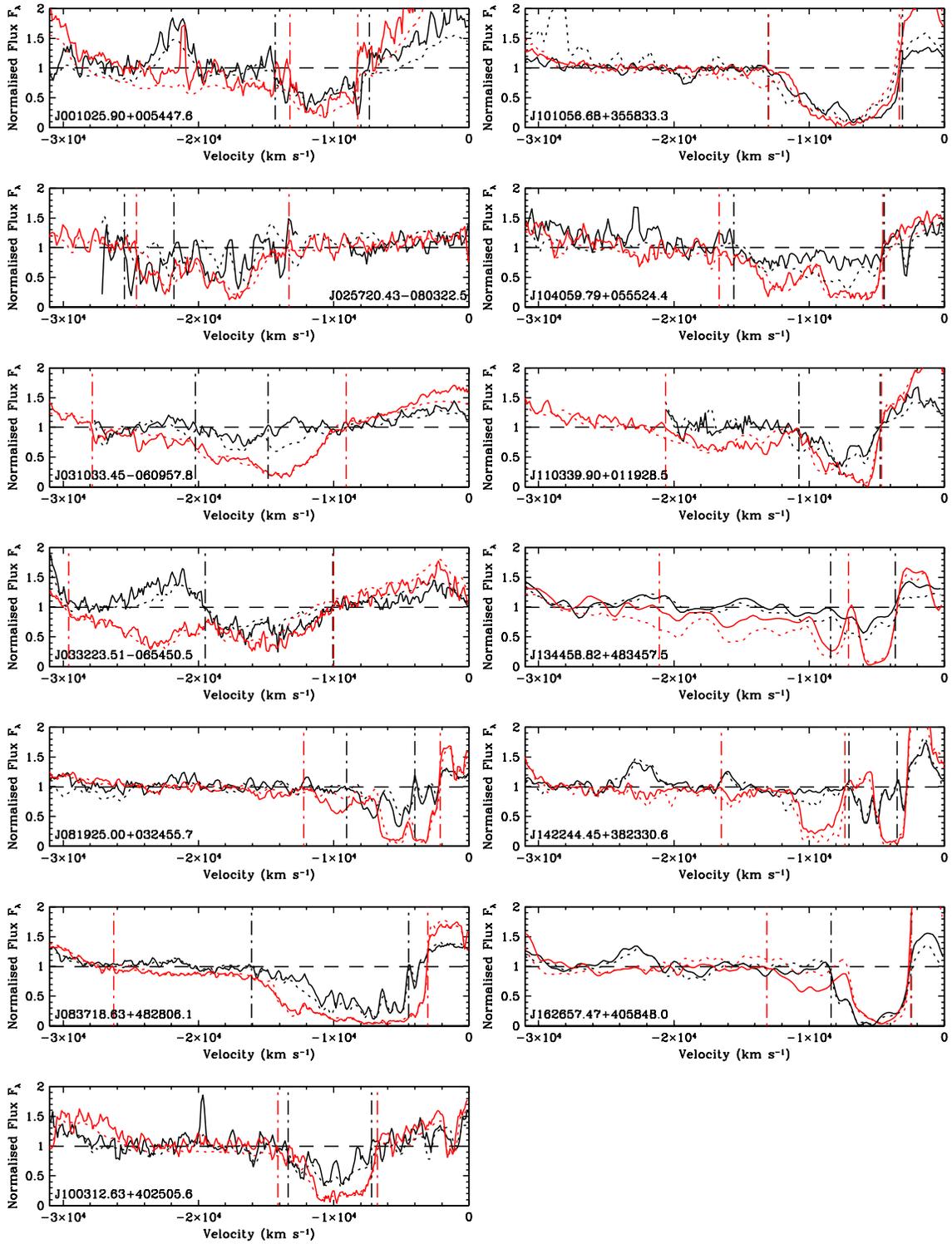}
\caption{Behaviour of Si~{\sc iv} (black) and C~{\sc iv} (red) BAL regions plotted in concurrent velocity space for each of the one ion variable, one ion non-variable quasar sample (from Table 5). Spectra are smoothed using a 3-pixel boxcar smoothing for clarity. Solid lines represent the SDSS spectra while dotted lines represent the Gemini/WHT spectra. Vertical dash-dot lines represent the BAL boundaries. A gap occurs between -12500 km s$^{-1}$ and -9200 km s$^{-1}$ in the SDSS spectrum of Si~{\sc iv} in J025720.43-080322.5 due to an unknown instrumental issue.}
\label{fig:varnonvar}
\end{figure*}

In previous work \citep{gibson10}, a significant correlation was found ($>$99$\%$) between the absolute fractional change in the total pseudo equivalent width in the absorption regions of C~{\sc iv} and Si~{\sc iv} (defined in that study as all spectral bins spanning 0 to -30000 km s$^{-1}$ from the rest wavelength of the red doublet component, as opposed to only the identified BAL regions used in this study) of 9 quasars over rest frame time intervals of $\approx$5-7 years. This high correlation depended on the exclusion of one outlying quasar, 1235+1453, which displayed a strong variability in the C~{\sc iv} BAL, but virtually no change in the Si~{\sc iv} BAL and showed a very blue continuum component which declines over subsequent observations and is not absorbed by the BAL region. A fitted line to this data (see Fig. 12 in \citet{gibson10}) gave a gradient of 1.44$\pm$0.45 and intercept -0.05$\pm$0.019 for the Si~{\sc iv} vs. C~{\sc iv} pseudo EW. Equivalent values from our data give a gradient of 1.88 and intercept of -0.11, well within the margin of error for both quantities.

	\subsection{Quasars containing both variable and non-variable BALs}

Of the 35 quasars containing at least one variable BAL, 15 also contained at least one non-variable BAL. Included in these are 8 containing a varying Si~{\sc iv} BAL and a non-varying C~{\sc iv} BAL, 5 containing a varying C~{\sc iv} BAL and a non-varying Si~{\sc iv} BAL and 5 containing a varying C~{\sc iv} BAL while a second C~{\sc iv} BAL did not vary. These instances are recorded in Table 5.

\begin{table*}
\begin{center}
\caption{Quasars containing both variable and non-variable BALs}
\begin{tabular}{l l l l}
\hline Object Name&Varying Si~{\sc iv}+non-varying C~{\sc iv}&Varying C~{\sc iv}+non-varying Si~{\sc iv}&Varying C~{\sc iv}+non-varying C{\sc iv}\\
\hline
SDSSJ001025.90+005447.6&no&yes&yes\\
SDSSJ004613.54+010425.7&no&no&yes\\
SDSSJ025720.43-080322.5&yes&no&no\\
SDSSJ031033.45-060957.8&yes&no&no\\
SDSSJ033223.51-065450.5&yes&no&no\\
SDSSJ081925.00+032455.7&yes&no&no\\
SDSSJ083718.63+482806.1&yes&no&no\\
SDSSJ100312.63+402505.6&yes&no&no\\
SDSSJ101056.68+355833.3&no&yes&no\\
SDSSJ104059.79+055524.4&yes&no&no\\
SDSSJ110339.90+011928.5&yes&no&no\\
SDSSJ134458.82+483457.5&no&yes&yes\\
SDSSJ142244.45+382330.6&no&yes&yes\\
SDSSJ162657.47+405848.0&no&yes&no\\
SDSSJ210436.62-070738.3&no&no&yes\\
\hline    
\end{tabular}
\label{tab:varnonvar}
\end{center} 
\end{table*}

The 13 showing variability in one ion and no variability in the other are particularly interesting as it is possible to investigate the extent, if any, of the velocity overlap. The spectral extent in velocity space of the BALs is illustrated in Fig. 8.

There is no velocity overlap at all between the two BALs in J142244.45+382330.6 (bottom right panel), while J134458.82+483457.5 (top left panel) and J025720.43-080322.5 (middle left panel) only show minimal overlap. The BALs in quasar J162657.47+405848.0 overlap but the region where the C~{\sc iv} BAL varies is largely non-overlapped. Of the 9 remaining quasars showing substantial velocity overlap in variable regions, 2 show C~{\sc iv} variation without Si~{\sc iv} variation while the remaining 7 showed the reverse situation. A lack of velocity space overlap in these objects suggests that their BALs are not part of the same outflow, see Section 6.2 for a further discussion.

	\subsection{BAL appearance and disappearance}

There is now a substantial body of evidence showing spectral regions containing BALs transforming into regions where absorption features no longer meet the BAL definition. This can occur for both HiBALs and
LoBALs \citep{hamann08,vivek12} and are referred to as cases of disappearing BALs. Disappearing BALs can transform into either spectral regions of no absorption or weaker absorption features. Such weaker features include mini-BALs \citep{hidalgo12,capellupo12} and NALs. Other BALs have been observed to emerge from apparently absorption free regions or from the types of aforementioned weaker absorption features and are referred to as appearing BALs. For this study, such events are confirmed only if the following two criteria are met, that disappearance or appearance takes place based on the BAL definition outlined in Section 4.1 and the significance meets the 2.5$\sigma$ variability criteria (also described in Section 4.1). Unlike in previous sections of this paper, we allow identification of BALs from second epoch data alone in order to record instances of BAL appearance. Table 6 provides a list of disappearing and appearing BALs along with their strengths as measured by the pseudo EW at epochs 1 and 2.

\begin{table*}
\begin{center}
\caption{Appearing (upper panel) and disappearing (lower panel) BALs. Total number of BALs in each quasar for each ion is given by T$_{no.}$. Pseudo equivalent widths are calculated over the velocity ranges which the single epoch BALs occupy.}
\begin{tabular}{l l l l l l}
\hline Object&Ion&$EW_{p1}$&$EW_{p2}$&$\Delta$$EW_{p}$&T$_{no.}$ (Si~{\sc iv}/C~{\sc iv})\\
 & &(km s$^{-1}$)&(km s$^{-1}$)&(km s$^{-1}$)& \\
\hline
SDSSJ100021.72+035116.5&Si~{\sc iv}&589$\pm$159&1509$\pm$99&920$\pm$187&2/2\\
SDSSJ100021.72+035116.5&Si~{\sc iv}&99$\pm$209&1229$\pm$123&1130$\pm$243&2/2\\
SDSSJ112733.69+343008.8&Si~{\sc iv}&1004$\pm$717&1549$\pm$295&545$\pm$775&1/2\\
SDSSJ112733.69+343008.8&C~{\sc iv}&0$\pm$177&787$\pm$52&787$\pm$184&1/2\\
\hline
SDSSJ085006.08+072959.0&C~{\sc iv}&1313$\pm$188&0$\pm$113&-1313$\pm$219&0/1\\
SDSSJ211718.17+010248.9&Si~{\sc iv}&1931$\pm$512&292$\pm$121&-1639$\pm$526&1/1\\
SDSSJ212412.60+095923.3&Si~{\sc iv}&1054$\pm$966&227$\pm$450&-827$\pm$1065&1/1\\
\hline
\end{tabular}
\label{tab:dapp}
\end{center}
\end{table*}

We note the appearance of a C~{\sc iv} BAL in the spectrum of J112733.69+343008.8 from a spectral region that was formerly almost entirely above continuum level (see Fig. 9). This case is noteworthy as it is accompanied by the emergence of a Si~{\sc iv} BAL where previously there had been a few non-contiguous NALs. This is more of a merging than a stregthening as the change in pseudo EW is consistent with zero. Although the width of the Si~{\sc iv} BAL is 16100 km s$^{-1}$, much wider than the 2600 km s$^{-1}$ C~{\sc iv} BAL, the deepest feature of the Si~{\sc iv} BAL largely occupies the same velocity range, while most of the Si~{\sc iv} BAL outside this range is so shallow as to be consistent with zero absorption. It is therefore likely that these two BALs represent absorption from the same outflow. In addition, this object hosts a second faster C~{\sc iv} BAL with a minimum velocity $\approx$1200 km s$^{-1}$ greater than that of the maximum velocity of the appearing C~{\sc iv} BAL. Although this velocity separation suggests that it is not part of the same outflow as the emerging C~{\sc iv} BAL, it also undergoes a strengthening of $\Delta$$EW_{p}$=480$\pm$437 km s$^{-1}$.

\begin{figure}
\centering
\resizebox{\hsize}{!}{\includegraphics[angle=0,width=8cm]{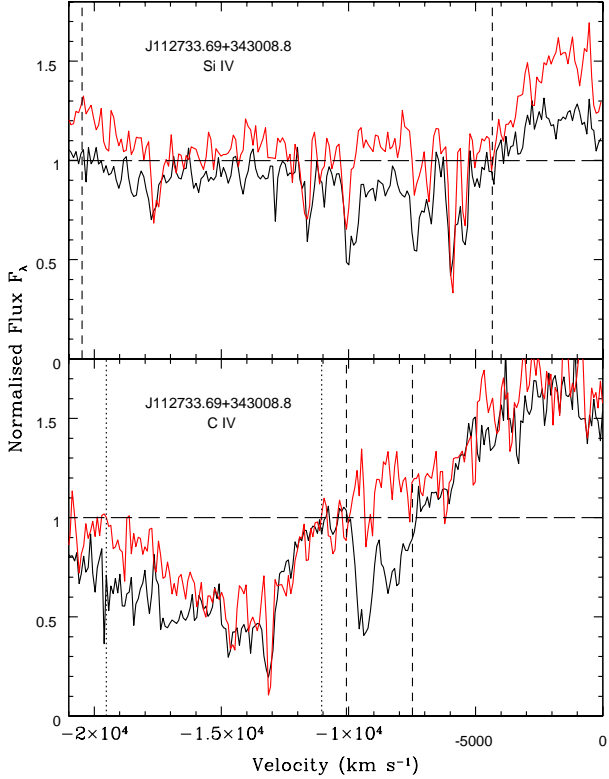}}
\caption{Appearance of a Si~{\sc iv} BAL (top panel) and a C~{\sc iv} BAL (bottom panel) in the quasar SDSSJ112733.69+343008.8 with SDSS observation in red and Gemini/GMOS observation in black. Vertical
dashed lines indicate boundaries of appearing BALs. Vertical dotted lines indicate the position of the second C~{\sc iv} BAL which undergoes strengthening.}
\label{fig:vels112733}
\end{figure}

In the case of J100021.72+035116.5 two Si~{\sc iv} BALs of similar strength appear to emerge in an almost contiguous absorption region save for a small break of $\approx$200 km s$^{-1}$ which protrudes
above the continuum and separates them (see Fig. 10). The lower velocity BAL of the two contained a mini-BAL at epoch 1 which strengthened significantly to become a BAL of width 3300 km~s$^{-1}$. The higher velocity BAL emerged from a spectral region having pseudo EW consistent with zero (99$\pm$209 km s$^{-1}$). At the second epoch the same spectral region contained a BAL of width 4000 km s$^{-1}$. The lower velocity Si~{\sc iv} BAL coexists with a C~{\sc iv} BAL at a mostly overlapping velocity range. Interestingly this C~{\sc iv} BAL does not show significant variability. There is also a second C~{\sc iv} BAL at higher velocity of width 9800 km s$^{-1}$ which overlaps most of the velocity range ($\approx$3000 km s$^{-1}$) of the higher velocity Si~{\sc iv} BAL. This C~{\sc iv} BAL also does not show significant variability. However only shallow parts of this BAL are in the overlapping region.

\begin{figure}
\centering
\resizebox{\hsize}{!}{\includegraphics[angle=0,width=8cm]{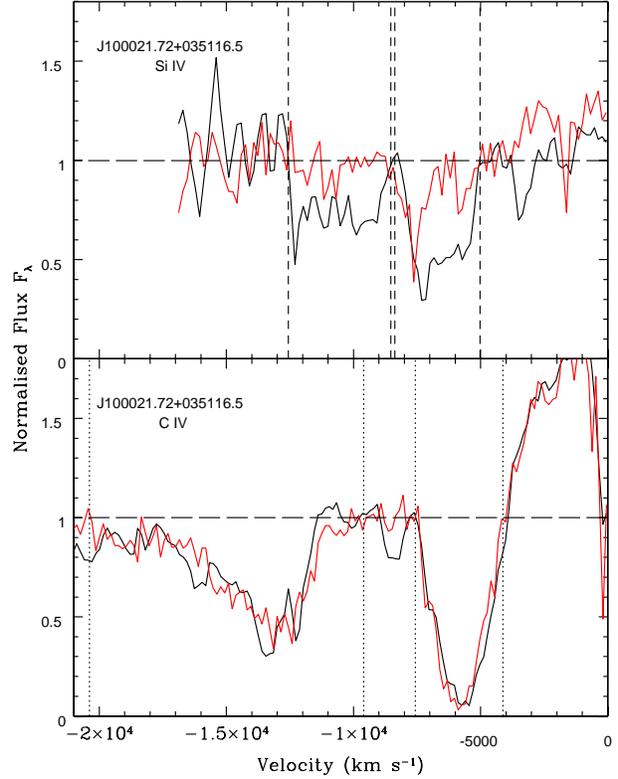}}
\caption{Appearance of a Si~{\sc iv} BAL (top panel) and a C~{\sc iv} BAL (bottom panel) in the quasar SDSSJ100021.72+035116.5 with SDSS observation in red and Gemini/GMOS observation in black. Vertical
dashed lines indicate boundaries of appearing BALs. Vertical dotted lines indicate the position of the C~{\sc iv} BALs which do not meet the variability criteria.}
\label{fig:vels100021}
\end{figure}

As the disappearing BALs all meet the variability criteria and are present in the first epoch, they are also listed in Table 2. Fig. 11 illustrates the spectra of these features across the two epochs for
all instances of BAL disappearance.

In \citet{filizak12}, 21 examples of C~{\sc iv} BAL disappearance were reported from a sample of 582 BALQSOs on similar rest frame timescales to this investigation (1.1 to 3.9 years). We observe one case of such behaviour of a C~{\sc iv} BAL from our sample, that of the disappearance from J085006.08+072959.0 where the BAL appears to have transformed into a narrower and significantly weaker feature ($\approx$4000 km s$^{-1}$ width without being sub-90$\%$ of continuum over a contiguous 2000 km s$^{-1}$) which meets the criteria for a mini-BAL. This changes the quasar's categorisation from BALQSO to non-BALQSO as there are no other BALs present for either ion in the second epoch. Over the entire velocity range formerly occupied by the BAL, the pseudo EW has declined to a level consistent with zero. The solitary C~{\sc iv} BAL disappearance in one quasar from our sample of 50 objects containing 59 C~{\sc iv} BALs gives a BAL disappearance rate of 1.7$\%$ and a fraction of quasars hosting C~{\sc iv} BALs undergoing an episode of disappearance of 2.5$\%$. This is consistent with the statistics reported in \citet{filizak12} of $\approx$2.3$\%$ of C~{\sc iv} BALs disappearing and $\approx$3.3$\%$ of BALQSOs hosting one of these disappearances over similar rest-frame timescales.

Disappearance of the Si~{\sc iv} BAL occurs in J211718.17+010248.9, with the remaining absorption reduced to a few narrow and weak absorption features. This is accompanied by the strong reduction in strength of the C~{\sc iv} line in this object by $\Delta$$EW_{p}$=-2684$\pm$382 km s$^{-1}$, with which it overlaps almost entirely in velocity range (this can be seen by comparing the spectra of J211718.17+010248.9 in Fig. 11 to Fig. 6). We also report the disappearance of a weak Si~{\sc iv} BAL in J212412.60+095923.3, reducing the total absorption in the region to a pseudo EW within 1$\sigma$ of zero. The weakness of this feature is highlighted by the fact that, as in the case of the appearing Si~{\sc iv} BAL in J112733.69+343008.8, the error on the change in pseudo EW makes it consistent with zero. Again this was accompanied by a strong reduction in the strength of the C~{\sc iv} feature of $\Delta$$EW_{p}$=-2427$\pm$673 km s$^{-1}$ over a very similar velocity range (again compare Fig. 11 with Fig. 6). The fraction of Si~{\sc iv} BALs that disappear is 2 out of 38, giving a disappearance rate of 5.26$\%$. This is also the disappearance rate per Si~{\sc iv} hosting quasar since no object hosts more than one Si~{\sc iv} BAL.

\begin{figure}
\centering
\resizebox{\hsize}{!}{\includegraphics[angle=0,width=8cm]{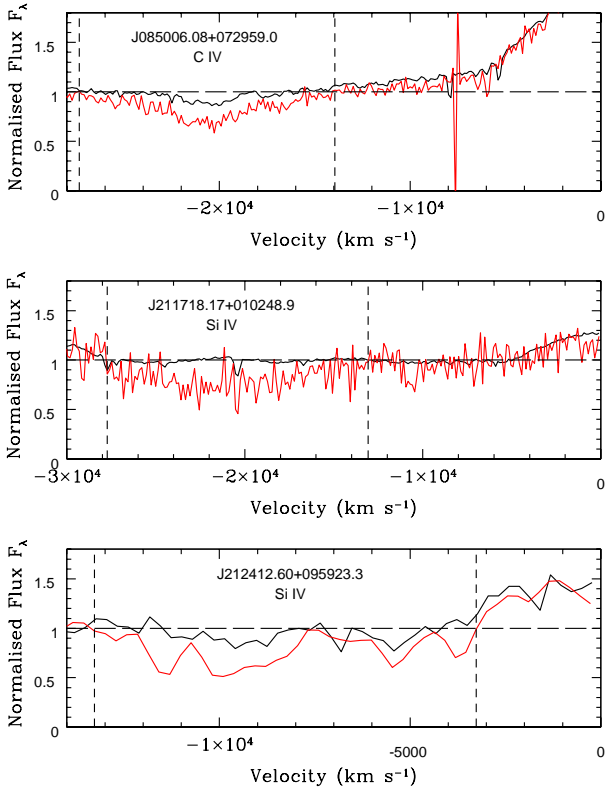}}
\caption{Disappearing BAL continuum normalised spectra in velocity space zeroed at emission line red doublet component. Epoch 1 spectra are in red, epoch 2 in black.}
\label{fig:figdisapp}
\end{figure}

	\subsection{Evidence of a covered continuum and uncovered BLR}

It was reported in \citet{arav99} that a section $\approx$1700 km s$^{-1}$ wide in the profile of the C~{\sc iv} BAL in FIRST J160354.2+300209 showed evidence of being due to an outflow which almost entirely absorbed the continuum while leaving the overlapping C~{\sc iv} broad emission line unabsorbed. This was thought to be due to an optically thick absorber which completely covers the continuum emitting region while leaving the much larger broad line region unobscured. We look for this same effect in our sample by seeking out those C~{\sc iv} troughs where the base overlaps with the emission line velocity range and shows a gradient consistent with the emission line profile. To create a model profile of the uncovered emission, an appropriate fraction of the continuum (up to 100$\%$) was selected to
be subtracted from the original reconstructed emission, whose value was determined by finding the fraction which minimised chi-square between the observed spectrum and this model over a velocity range where the trough profile appeared to be similar to the continuum absorbed emission profile. A good match was found in 5 C~{\sc iv} BALs, one in each of 5 quasars. Details for these C~{\sc iv} BALs is provided in Table 7, while their spectra are illustrated in Fig. 12. None of these objects showed evidence for the same effect in Si~{\sc iv} BALs.

\begin{table*}
\begin{center}
\caption{C~{\sc iv} BALs showing evidence of both an absorbed continuum and unabsorbed broad line region.}
\begin{tabular}{l l l l l}
\hline Object&V$_{max}$&V$_{min}$&Epoch 1 covering fraction ($\pm$10$\%$)&Epoch 2 covering fraction ($\pm$10$\%$)\\
 &(km s$^{-1}$)&(km s$^{-1}$)& & \\
\hline
SDSSJ035335.67-061802.5&-12600&-11800&0.71&0.86\\
SDSSJ073535.44+374450.4&-12900&-8100&0.78&0.82\\
SDSSJ100312.63+402505.6&-10500&-8000&0.84&0.88\\
SDSSJ114722.09+373720.7&-14300&-8400&0.64&0.71\\
SDSSJ143632.25+501403.6&-12000&-8700&0.79&0.88\\
\hline
\end{tabular}
\label{tab:belolap}
\end{center}
\end{table*}

When considering the variation between the two epochs, we find no strong evidence for a change in the covering fraction over these velocity ranges. It is also notable that 4 of the 5 C~{\sc iv} BALs are not variable. Only the BAL in SDSSJ073535.44+374450.4 meets the variability criteria, however within the velocity range over which the BAL profile follows the emission line profile, this feature also shows
no variability.

\begin{figure*}
\centering
\includegraphics[width=170mm]{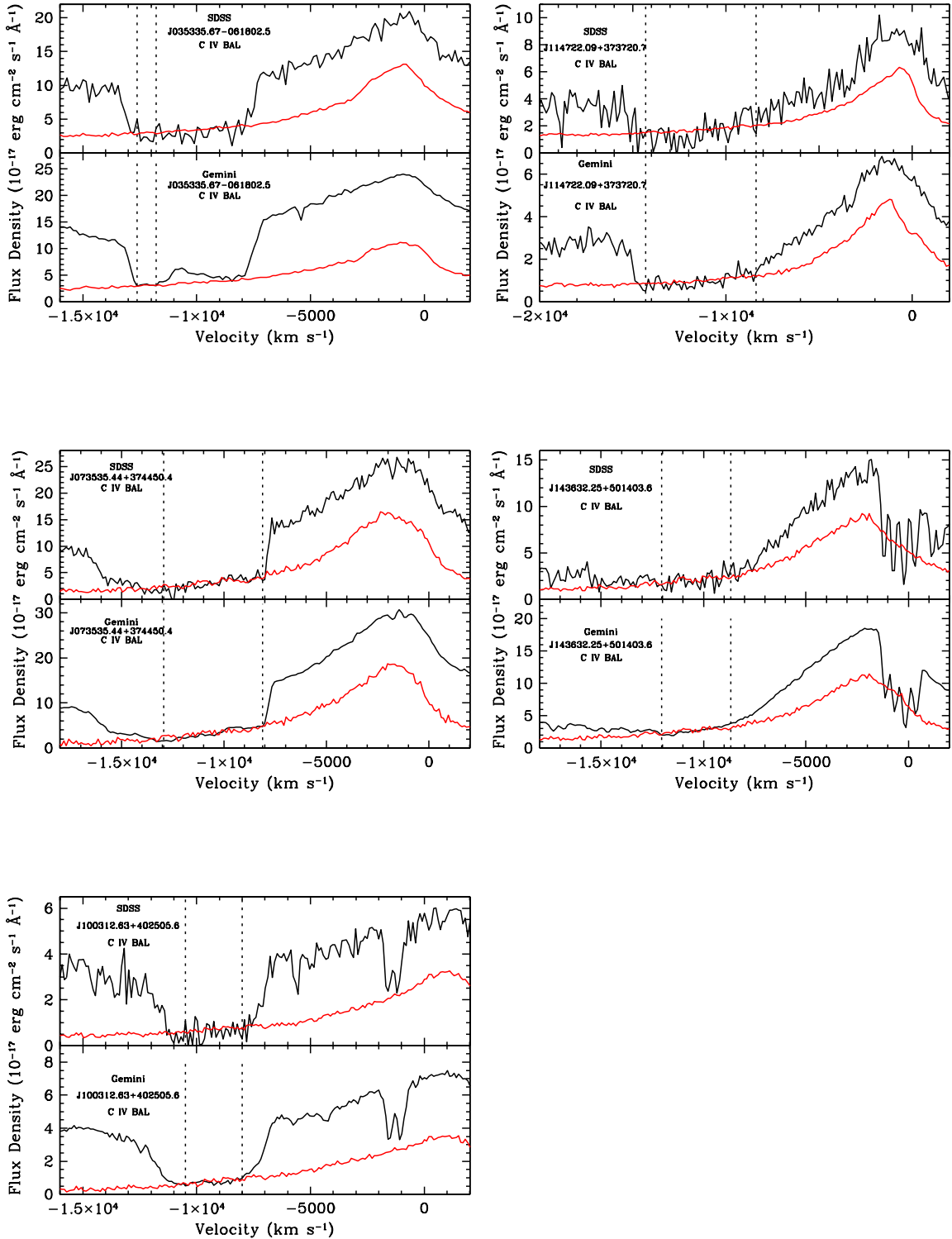}
\caption{SDSS observations (upper panels) and Gemini/GMOS observations (lower panels) for 5 C~{\sc iv} BALs showing evidence for a gas geometry where the broad emission line is unobscured while most of the continuum region is covered by an optically thick absorber. Reconstruction minus covered continuum is in red, while observed spectrum is in black.}
\label{fig:belplots}
\end{figure*}

	\subsection{Other notable objects}

	\subsubsection{SDSSJ113831.42+351725.3}

This object contains one C~{\sc iv} and one Si~{\sc iv} BAL which show the largest absolute change in pseudo equivalent out of any BAL of their respective ions in the entire variable BAL sample. The C~{\sc iv} BAL undergoes a change of $\Delta$EW$_{p}$=-4142$\pm$817, larger than the corresponding absolute change in Si~{\sc iv} of $\Delta$EW$_{p}$=-2880$\pm$647 km s$^{-1}$. However the situation is reversed when considering the fractional change in pseudo EW, which the Si~{\sc iv} BAL has a value of $\Delta$$EW_{p}$$/$$\langle$$EW_{p}$$\rangle$=-0.979$\pm$0.245, while for the C~{\sc iv} BAL it is $\Delta$$EW_{p}$$/$$\langle$$EW_{p}$$\rangle$=-0.606$\pm$0.125. Apart from variability, the properties of this quasar are not exceptional.

	\subsubsection{SDSSJ164152.30+305851.7}

This is the only quasar in our sample showing evidence of LoBAL absorption, with both Mg~{\sc ii} and Al~{\sc iii} BALs visible in both the WHT and Gemini Spectra. It is likely that a C~{\sc ii} BAL overlaps the Si~{\sc iv} broad absorption in this quasar as otherwise the Si {\sc iv} BAL would have extreme properties, with a pseudo EW of $\langle$EW$_{p}$$\rangle$=26389$\pm$742 km s$^{-1}$, a width of 28661$\pm$512 km s$^{-1}$ and a depth of $\langle$B$_{depth}$$\rangle$=0.909$\pm$0.029. These quantities would be by far the largest for any Si~{\sc iv} BAL in the variable sample (the second strongest Si~{\sc iv} BAL sample has $\langle$EW$_{p}$$\rangle$=7255$\pm$214 km s$^{-1}$) and would also be stronger than any C~{\sc iv} BAL in the variable sample, including the C~{\sc iv} BAL in the same object. This would be unusual as Si~{\sc iv} BALs are weaker than C~{\sc iv} BALs in every other object containing variable BALs of both kinds. Aside from the presence of LoBAL absorption, this quasar is also interesting as it exhibits a C~{\sc iv} BAL with the highest pseudo EW of any C~{\sc iv} BAL in the variable sample ($\langle$EW$_{p}$$\rangle$=13526$\pm$593 km s$^{-1}$).

	\section{Conclusions}

	\subsection{Epoch separation time and velocity dependence}

By performing a comparative study using data sets from \citet{barlow93}, L07 and their own sample, G08 found an increase in C~{\sc iv} absorption variability with increased epoch separation time. We find a mean value of $|$$\Delta$$EW_{p}$$/$$\langle$$EW_{p}$$\rangle$$|$=0.43$\pm$0.03 for all C~{\sc iv} BALs identified as variable. This compares with a mean value of $|$$\Delta$$EW_{p}$$/$$\langle$$EW_{p}$$\rangle$$|$=0.55$\pm$0.03 for the sample of 16 BALs showing significant variability in L07. However their mean is skewed by three outlying features, the lowest of which has an EW more than double that of the next most variable BAL. If these BALs are excluded, a mean value of $|$$\Delta$$EW_{p}$$/$$\langle$$EW_{p}$$\rangle$$|$=0.31$\pm$0.03 is recovered, lower than the value for our sample and consistent
with a positive correlation between BAL variability amplitude and epoch separation time when considering timescales ranging from months to several years. However, the range of timescales studied within our quasar sample are not sufficient to provide evidence of variability dependence on our epoch separation time.

We find no evidence for the proposed spike in variability in the velocity range 12000 to 15000 km s$^{-1}$ proposed in L07. In fact, no trends are found relating the velocity width or BAL centroid velocity with variability. If variability is mainly driven by differences in covering fraction, then the absence of any correlation between BAL
centroid velocity and variability amplitude could indicate that the outflow velocity along the line of sight is not well correlated with velocity across the line of sight.

	\subsection{What gives rise to BAL variability?}

Fractional pseudo equivalent width variability is strongly inversely correlated with the mean BAL depth averaged over the two epochs. This is what would be expected if ionizing continuum changes altering the fractions of the ions making up the outflowing gas were responsible for variability in the strength of the BALs, as deeper troughs may indicate an optically deeper outflow. However this is questionable for two reasons: (1) the depth measure and the epoch averaged EW show a strong positive correlation, suggesting that this relationship is an artifact of the denominator in the fractional change, and (2) we find no strongly significant correlation between rest frame i-band luminosity (M$_{i}$) and BAL variability. There is a moderately significant probability of an inverse correlation between absolute change in pseudo EW and absolute i-Band luminosity in the variable C~{\sc iv} BAL sample, however under the assumption of lower luminosity sources driving stronger variability in the case where ionization changes are responsible, it would be expected that we would see a corresponding signal indicating absolute change in pseudo EW is inversely correlated with $\langle$B$_{depth}$$\rangle$ for C~{\sc iv} BALs, which is not found. In fact, Si~{\sc iv} BALs show a moderate probability of the same correlation, leading to the conclusion that there are no real correlations in these cases and that there is no strong evidence for ionization changes driving BAL variability.

An alternative explanation is that the observed variability is driven by covering fraction changes. One such possibility is an inhomogeneous absorber model in which the column density along the line of sight is due to clumpy absorbers which can move laterally into and out of the line of sight at any radial distance from the continuum source. The clumpy inhomogeneous model has received support from \citet{capellupo12} who found the strongest variability within small velocity regions of BAL troughs, which may correspond to movements of individual clouds with narrow velocity ranges. An alternative to the
inhomogeneous absorber model is a so-called pure partial coverage scenario in which the gas is relatively homogeneous along the radial axis but moves laterally with respect to the line between the Earth
and the quasar emission region.

The Si~{\sc iv} BALs are observed to vary more often than the C~{\sc iv} BALs and within those quasars that show both types of BAL varying, the Si~{\sc iv} BALs undergo the greatest fractional variability. This evidence would seem to favour the clumpy model over the homogeneous scenario as it could be due to an outflow in which the greater column density of C~{\sc iv} ions over Si~{\sc iv} means that C~{\sc iv} is more likely to be saturated. In such an outflow, a cloud moving into the line of sight may not affect the absorption line profile since a particular cloud would be entering into a portion of the outflowing column which is already opaque at the narrow range of wavelengths that the cloud could absorb. This could result in a scenario where a column of clumpy absorbers only partially covers the emission from the quasar, but due to the spatial extent along the line of sight of the entire outflow, a cloud moving into the line of sight does not affect the absorption profile. Higher frequency of saturation of C~{\sc iv} as compared to Si~{\sc iv} (due to the greater abundance of carbon compared to silicon, or possibly ionization effects favouring C~{\sc iv} over Si~{\sc iv}) is also supported by the fact that of the five quasars showing BALs with an uncovered BLR and covered continuum, the uncovered BLR emission spectra do not fit the Si~{\sc iv} BAL profiles. In such a saturated C~{\sc iv} BAL case, moving clouds into and out of the line of sight would not significantly change the depth of C~{\sc iv} troughs when compared with Si~{\sc iv} variability.

The fact that the 14 quasars showing variability in both ions have BALs with largely overlapping outflow velocities, in which the direction of change is the same, is strong evidence for the two types of BAL being part of the same outflow and influenced by the same variability mechanism in these cases. The sample of 9 quasars showing one type of BAL varying while the other does not, with substantial velocity overlap between the two, seems to contradict such a conclusion. However, the variable BAL in 7 of these cases is Si~{\sc iv}, while the C~{\sc iv} BAL does not vary. If C~{\sc iv} is saturated in these cases, this would allow Si~{\sc iv} to vary without affecting the C~{\sc iv} BAL variability amplitude in the overlapping velocity range, allowing consistency with the clumpy absorber model.

Despite this, ionization changes are not entirely ruled out as a driver of BAL variability. The luminosity dependence that was used as a proxy for EUV variability is two steps removed from a direct observation of the ionizing continuum flux and it is possible that the complexity of the response of the outflowing gas to ionizing continuum changes may not be analogous to that of the emission lines. The hint of a positive correlation between the fractional variability of Si~{\sc iv} and C~{\sc iv} where an approximately linear relationship exists between the two quantities with a gradient not equal to 1 (see section 5.1) would seem to fit more naturally with ionizing continuum variability as the explanation rather than gas line of sight changes. However this is far from certain given the value of only
98$\%$ correlation probability derived from Spearman rank test-and the fact that the least squares fit line only appears to fit 8 out of the 14 quasars showing variability in both types of BAL. Other possible indicators of ionizing continuum changes include 2 cases of C~{\sc iv} variability which are {\em unaccompanied\/} by Si~{\sc iv} variability at the same velocities. This is hard to explain in terms of gas covering fraction changes as it would require some sort of spatial separation between Si~{\sc iv} and C~{\sc iv} outflows. Calculations of theoretical ionization fractions across ionization parameter space suggest this is unlikely as there is substantial overlap in the Log Ion fraction vs. Log ionization parameter curves of each ion (see Fig. 4 in \citet{hamann01}).

Future investigations may be able to find more definitive evidence for continuum variations by accurately measuring continua in a high S/N quasar sample across several epochs with photometric quality spectra and looking for correlations with BAL variability. An underlying assumption in investigations into continuum variability correlations with BAL variability is that the outflowing gas is dense enough to undergo recombinations on relatively short timescales. Although the column densities of absorbers can be estimated using optical depth calculations (these are lower limits in the case of non-black
saturation), the extent of the BAL flow is still a subject of debate, with estimates placing the flow as far as the kiloparsec scale from the central engine \citep{dunn10}. If this was the case, given the
estimated outflow column densities of logN$_{H}$$\approx$21 to 22~cm$^{-2}$ \citep{hamann08} the gas density could be as low as $\approx$10 cm$^{-3}$. Given a recombination coefficient value of $\approx$10$^{-11}$ cm$^3$ s$^{-1}$ for C~{\sc iv} and Si~{\sc iv} and assuming the gas is almost completely ionized, recombination timescales could be as long as a few hundred years. Attempts to find correlations between luminosity or continuum variability and BAL variability would in this case give no insight into how ionizing continuum changes affect the outflowing gas.

In order to gain a greater insight in to the physical influences driving BAL variability, larger BALQSO sample sizes are needed. The ongoing Baryon Oscillation Spectroscopic Survey (BOSS), part of the Sloan Digital Sky Survey-III (SDSS-III) \citep{eisenstein11} should prove helpful in this regard. One of the ancillary projects conducted as part of this survey will provide much greater numbers of BALQSO
repeat observations on multi-year timescales than previous studies by comparing BOSS BALQSO spectra with earlier SDSS-I and II data \citep{filizak12}.

	\subsection{BAL disappearance and BAL lifetimes}

The fraction of C~{\sc iv} BALs in our sample which disappear is 1.7$\%$, lower than the 5.3$\%$ value found for Si~{\sc iv} BALs. However this results from only three examples of BAL disappearance, so definitive conclusions are not drawn, except to say that the disappearance rate for C~{\sc iv} is broadly consistent with that found in \citet{filizak12}. If Si~{\sc iv} BALs do show a greater tendency to disappear, this would fit in with the general picture of Si~{\sc iv} BALs exhibiting greater occurrences of variability and greater amplitudes of variability in a variable BAL sample. The disappearing BALs are weaker than average, as are the BALs which appear. In fact the disappearing C~{\sc iv} BAL is the weakest of all the C~{\sc iv} BALs in the variable sample. This conforms with the tendency of shallower BALs to exhibit stronger fractional variability.

The average lifetime of a BAL can be estimated by dividing the fraction of BALs that disappear by the average rest-frame lifetime of the observations. This gives an average BAL rest frame lifetime of $\approx$142 years for C~{\sc iv} and $\approx$43 years for Si~{\sc iv} based on the timescales and disappearing fractions from our sample. Our estimated C~{\sc iv} lifetime is similar to the value of
109$^{+31}_{-22}$ years estimated in \citep{filizak12}.

	\subsection{Outflow geometry insight provided by covered-continuum-uncovered-BLR subsample} 

Reverberation studies of the variable continuum and broad emission-line region in AGN, indicate that the ``size'' of the BLR scales with UV ionizing continuum luminosity such that $R\propto L_{UV}^{1/2}$ \citep{Bentz06}, as predicted by photoionization models. The luminosities of the five objects showing evidence of C~{\sc iv} BALs shaped by a covered continuum+uncovered BLR are not biased towards more or less luminous objects. This indicates that the difference between this sample and other BAL quasars is not due to a smaller BLR and is instead determined by the geometry of the outflow along the line of
sight. Given that the shape of the troughs in this subsample are well modeled by a mostly absorbed continuum and an entirely unobscured BLR, the maximum characteristic radius of any clumpy absorber clouds would have to be similar to the UV continuum emitting region size. Assuming that the central supermassive black hole is accreting at close to the Eddington rate and a peak ionizing UV continuum temperature
corresponding to a photon energy of $\approx$50 eV, the ratio of the UV luminosity to the bolometric luminosity will be of the order $\approx$10$^{-1}$ \citep{kelly08}. Given the bolometric luminosity
span of these 5 objects (obtained from \citet{shen11}) the size of the UV continuum emitting region and hence the maximum size of an obscuring cloud will be in the range $\approx$10$^{13}$ to 10$^{14}$ cm. Two possibilities therefore exist, either the characteristic size of the clouds or the radius (across the line of sight) of a homogeneous outflow column must be able to vary greatly between objects, from continuum region size up to $\approx$R$_{BLR}$, or the radius of the absorbing column emerging from the disc can vary between quasars by having considerably different numbers of approximately UV continuum emitting region sized clouds across the sight line.

	\section{Summary}

A brief summary of the main results is provided here.\\
\\
\noindent 1. Si~{\sc iv} BALs are more likely to be variable than C~{\sc iv} BALs (58$\%$ vs. 46$\%$ respectively). Within the sample of variable BALs, the average fractional change in pseudo EW is higher
for Si~{\sc iv} BALs (0.577 vs. 0.367 for C~{\sc iv} BALs when excluding the LoBAL quasar SDSSJ164152.30+305851.7).

\noindent 2. Both Si~{\sc iv} and C~{\sc iv} BALs have fractional changes in pseudo EW which are strongly correlated (greater than 99.9$\%$ significance) with their epoch averaged depths. We suggest that this is a result of the strong positive correlation between BAL depth and pseudo EW rather than a real physical effect. No strong evidence is found of absolute change in pseudo EW dependance on epoch averaged pseudo EW, nor any correlation of absolute change in pseudo EW with BAL depth or width.

\noindent 3. No correlation is found between BAL variability and the absolute i-band luminosity of their host quasars. Since ionizing continuum changes drive broad emission line variability, this result
suggests that variability of the ionizing continuum does not play a significant role in BAL variability, assuming the outflowing gas is in ionization equilibrium with the photoionizing continuum.

\noindent 4. Within each of the 14 quasars exhibiting both a varying Si~{\sc iv} and a varying C~{\sc iv} BAL, the change in pseudo EW is in the same direction (strengthening or weakening) across the two
BALs. Furthermore, each of these objects show significant overlap in velocity space of the spectral regions spanning the two, suggesting that the BALs of both ions originate in the same outflow. There is marginal evidence (98$\%$ significance) that the fractional change in pseudo EW of Si~{\sc iv} and C~{\sc iv} BALs are correlated.

\noindent 5. Examples of both BAL appearance and disappearance are found, with instances in both Si~{\sc iv} (3 appearance, 2 disappearance) and C~{\sc iv} (1 appearance, 1 disappearance). Their relative recurrence rates suggest Si~{\sc iv} and C~{\sc iv} BAL lifetimes of 43 years and 142 years respectively. Appearing and disappearing BALs are weaker than average, complying with the general trend noted in point 2.

\noindent 6. There is evidence in 5 objects of a mostly covered continuum region combined with an uncovered BLR. This can be seen as a trough profile in deep C~{\sc iv} BALs which follows the emission line
profile once a substantial fraction of the continuum has been subtracted. This suggests a range in maximum cloud radius for the clumpy inhomogeneous absorber model of $\approx$10$^{13}$ to 10$^{14}$ cm for the outflows from these quasars, assuming accretion is close to the Eddington rate.

\section*{Acknowledgements}

This work is supported at the University of Leicester by the Science and Technology Facilities Council (STFC) and is partly based on observations obtained at the Gemini Observatory, which is operated by
the Association Research in Astronomy, Inc., under a cooperative agreement with the NSF on behalf of the Gemini partnership. It is also based on observations made with the WHT/ISIS operated on the island of
La Palma by the Isaac Newton Group in the Spanish Observatorio del Roque de los Muchachos of the Instituto de Astrofísica de Canarias, and spectroscopic observations from Data release 6 of the Sloan Digital Sky Survey (SDSS). We wish to thank our anonymous referee for their useful comments and thorough review of the draft. JTA acknowledges the award of an ARC Super Science Fellowship.

Funding for the SDSS and SDSS-II has been provided by the Alfred P. Sloan Foundation, the Participating Institutions, the National Science Foundation, the U.S. Department of Energy, the National Aeronautics and Space Administration, the Japanese Monbukagakusho, the Max Planck Society, and the Higher Education Funding Council for England. The SDSS Web Site is http://www.sdss.org/.

The SDSS is managed by the Astrophysical Research Consortium for the Participating Institutions. The Participating Institutions are the American Museum of Natural History, Astrophysical Institute Potsdam, University of Basel, University of Cambridge, Case Western Reserve University, University of Chicago, Drexel University, Fermilab, the Institute for Advanced Study, the Japan Participation Group, Johns Hopkins University, the Joint Institute for Nuclear Astrophysics, the Kavli Institute for Particle Astrophysics and Cosmology, the Korean Scientist Group, the Chinese Academy of Sciences (LAMOST), Los Alamos National Laboratory, the Max-Planck-Institute for Astronomy (MPIA), the Max-Planck-Institute for Astrophysics (MPA), New Mexico State University, Ohio State University, University of Pittsburgh, University of Portsmouth, Princeton University, the United States Naval Observatory, and the University of Washington.


\bibliography{wildyrevd}

\bibliographystyle{mn2e}

\bsp

\label{lastpage}

\end{document}